\documentclass{jfm}
\usepackage{graphicx}
\usepackage{epstopdf, epsfig}
\usepackage{color}
\usepackage{subfigure}
\usepackage{amsmath}
\usepackage{multirow}
\usepackage{mathabx}
\usepackage{verbatim}
\usepackage{lineno}
\usepackage{tikz}
\usepackage[colorlinks=true, linkcolor=blue, urlcolor=blue, citecolor=blue]{hyperref}

\newcommand{\mkf}{\textcolor{black}}
\definecolor{Gr50}{rgb}{0.082, 0.491, 0.224}
\newcommand{\dc}{\textcolor{black}}

\title{Turbulent drag reduction by spanwise wall forcing. Part 2: High-Reynolds-number experiments.}
\author{D. Chandran$^1$, A. Zampiron$^2$, A. Rouhi$^{3}$\corresp{\email{amirreza.rouhi@ntu.ac.uk}}, M. K. Fu$^{4}$, D. Wine$^5$\\ B. Holloway$^5$,
A. J. Smits$^6$ \and I. Marusic$^1$}
\affiliation{
$^1$Department of Mechanical Engineering,
University of Melbourne, Victoria 3010, Australia\\[5pt]
$^2$ School of Engineering, University of Aberdeen, Aberdeen, UK\\[5PT]
$^3$Department of Engineering, Nottingham Trent University, Nottingham, UK\\[5pt]
$^4$Graduate Aerospace Laboratories (GALCIT),
Caltech, Pasadena, CA, USA\\ [5pt]
$^5$Intellectual Ventures, Bellevue, WA, USA\\[5pt]
$^6$Department of Mechanical and Aerospace Engineering, Princeton University, Princeton, NJ, USA}

\shorttitle{Turbulent drag reduction by spanwise wall forcing. Part 2: Experiments}

\begin{document}

\maketitle

\begin{abstract}
Here, we present measurements of \dc{turbulent drag reduction in boundary layers} at high friction Reynolds numbers in the range of $4500 \le Re_\tau \le 15000$. The efficacy of the approach, using streamwise travelling waves of spanwise wall oscillations, is studied for two actuation regimes: (i) inner-scaled actuation (ISA), as investigated in Part 1 of this study, which targets the relatively high-frequency structures of the near-wall cycle, and (ii) outer-scaled actuation (OSA), which was recently presented by \citeauthor{marusic2021nature} (\textit{Nat.\ Commun.}, vol.\ 12, 2021) for high-$Re_\tau$ flows, targeting the lower-frequency, outer-scale motions. 
Multiple experimental techniques were used, including a floating-element balance to directly measure the skin-friction drag force, hot-wire anemometry to acquire long-time fluctuating velocity and wall-shear stress, and stereoscopic-PIV (particle image velocimetry) to measure the turbulence statistics of all three velocity components across the boundary layer. Under the ISA pathway, drag reduction of up to 25\% was achieved, but mostly with net power saving losses due to the high-input power cost associated with the high-frequency actuation. The low-frequency OSA pathway, however, with its lower input power requirements, was found to consistently result in positive net power savings of 5 - 10\%, for moderate drag reductions of 5 - 15\%. The results suggest that OSA is an attractive pathway for energy-efficient drag reduction in high Reynolds number applications. Both ISA and OSA strategies are found to produce complex inter-scale interactions, leading to attenuation of the turbulent fluctuations across the boundary layer for a broad range of length and time scales.
\end{abstract}
 


\section{Introduction}
\label{sec:Intro}
\begin{figure}
	\begin{center}
		\includegraphics[width=1\linewidth,trim = 0mm 0mm 0mm 0mm, clip]{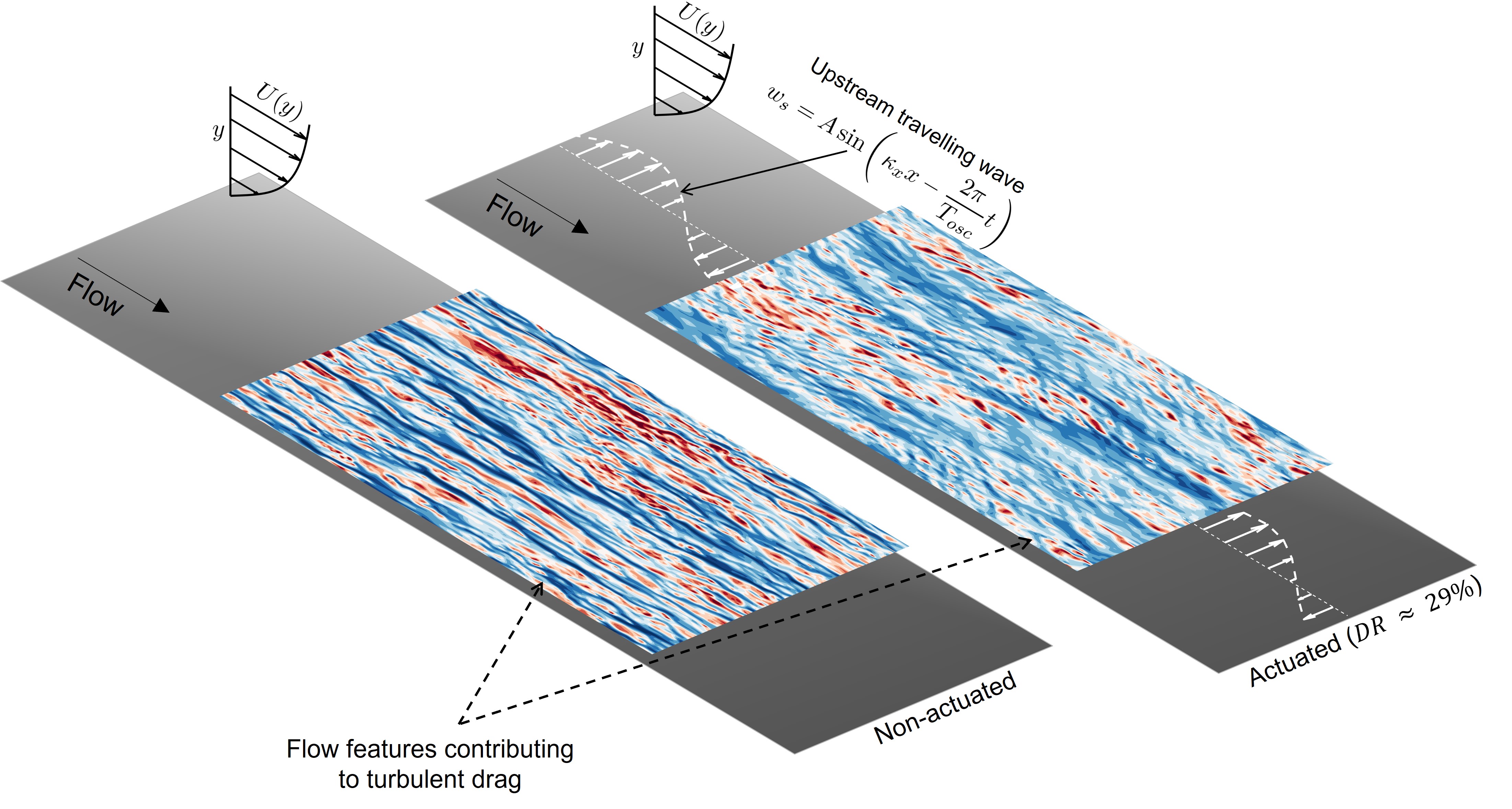}
		\caption{Visualization of near-wall flow features for the non-actuated case and an actuated case with a streamwise travelling wave of spanwise velocity. The time-scale of actuation is $T_{osc}^+ = 140$, resulting in a drag reduction of 29\%.}
		\label{fig:LES_illus}
	\end{center}
\end{figure}

The sensitivity of wall-bounded turbulent flows to boundary conditions means that relatively modest adjustments made to the wall can significantly modify fluid behaviour, including drag. This concept has driven the development of a wide variety of flow control methods (see, for example, \citealp{kim2011physics}). One such method is the active flow control strategy of imposing streamwise travelling waves of spanwise velocity at the wall \citep{quadrio2009streamwise}, which is the focus of the current study and is briefly reviewed in Part 1 \citep{rouhi2022part1}. A more extensive review is given by \cite{ricco2021review}. Here, the forcing arises due to an imposed surface movement with
 \begin{equation}
	w_s(x,t)=A \sin{\bigg(\kappa_x x - \frac{2\pi}{T_{osc}} t \bigg),}
	\label{wallmotion}
\end{equation}
where $w_s$ is the instantaneous spanwise velocity at the wall, $A$ and $T_{osc}$ are the amplitude and time period of spanwise actuation, respectively, and $\kappa_x = 2\pi/\lambda$ is the streamwise wavenumber of the travelling wave; $\lambda$ is the wavelength. In Part 1, \eqref{wallmotion} was presented based on the angular frequency of actuation $\omega = 2\pi/T_{osc}$. Streamwise, wall-normal and spanwise coordinates are denoted by $x$, $y$, and $z$, respectively (with corresponding instantaneous velocities $u$, $v$ and $w$), and $t$ is time. A schematic of this forcing is shown on the right side of figure \ref{fig:LES_illus}. 

Numerical studies indicate that this approach shows great promise \citep{quadrio2011drag}, but applying this control strategy in practice has remained a challenge. This is especially due to the difficulties in accurately reproducing, predicting, or modelling the turbulent dynamics that are encountered in high Reynolds number practical flows. Previous experimental \citep{auteri2010experimental,bird2018experimental} and numerical investigations (\citealp{quadrio2009streamwise,hurst2014effect,gatti2016reynolds,gatti2018spectra,skote2022drag} etc.) of this control strategy have been restricted to lower Reynolds numbers, $Re_\tau \le 1000$. Here, the friction Reynolds number is defined as $Re_\tau = \delta u_{\tau_0}/\nu$, where, $u_{\tau_0}=\sqrt{\tau_{w_0}/\rho}$ is the friction velocity, $\tau_{w_0}$ is the wall-shear stress, $\delta$ is the boundary layer thickness \dc{of the non-actuated flow}, $\rho$ is the fluid density and $\nu$ is the fluid kinematic viscosity. The subscript `$0$' indicates parameters evaluated at the non-actuated reference condition.  
The superscript `$+$' indicates normalization using viscous length ($\nu/u_{\tau_0}$) and velocity ($u_{\tau_0}$) scales. The superscript `$*$' indicates the normalization using viscous scales where the actual friction velocity is considered, i.e., $u_\tau$ of the drag-altered flow for the actuated cases. 

At relatively low Reynolds numbers, $Re_\tau = \mathcal{O}(10^2)$ to $\mathcal{O}(10^3)$, near-wall streaks are the statistically dominant turbulent structures close to the wall and follow inner scaling \citep{Kline1967,Smits2011a}. That is, their features scale with velocity $u_{\tau_0}$ and length $\nu/u_{\tau_0}$. The predominant time-scale of the near-wall streaks is found to be $T^+ = T u_{\tau_0}^2/\nu \approx 100$ and their characteristic streamwise and spanwise lengths are $1000\nu/u_{\tau_0}$ and $100\nu/u_{\tau_0}$, respectively. Flow control schemes that are implemented at the wall often prescribe a forcing of similar time and/or length scales to couple with these features and achieve the best result \citep{jung1992suppression,choi1998drag,quadrio2004critical,choi2011plasma,kim2011physics,tomiyama2013direct,skote2013,lozano2020non}. 

The actuation strategy of targeting the near-wall streaks, which we refer to as {\em inner-scaled actuation} (ISA) was the focus of Part 1 of this study where large-eddy simulations (LES) were used. 
Figure \ref{fig:LES_illus} shows a visualization of the near-wall flow field from LES at $Re_\tau = 951$ for the non-actuated case and an actuated case where the drag reduction $DR \approx 29\%$.  Here, $DR= 1-\overline{\tau_w}/\overline{\tau_{w_0}}$, where $\overline{\tau_{w}}=\rho u_\tau^2$ is the time-averaged wall shear stress for the actuated case.   
The actuation is seen to deplete the streaks and attenuate the intensity of turbulent fluctuations near the wall. Although the performance of ISA is reported to deteriorate with increasing Reynolds number, it is still observed to yield significant drag reduction of $DR \approx 25\%$ at $Re_\tau = 4000$ (Part 1) and $Re_\tau = 6000$ \citep{marusic2021nature}. Unfortunately, however, targeting near-wall streaks typically implies high-frequency actuation, and thus high-input power requirements, so that ISA often ends up being energy inefficient (\citealp{marusic2021nature}).
 
An alternate, more energy-efficient, pathway to drag reduction at high Reynolds numbers targets the larger scale, outer-region, structures \citep{marusic2021nature}. We refer to this relatively lower-frequency actuation strategy as {\em outer-scaled actuation} (OSA). By outer-scaled, we refer to all motions that scale with $y$ and/or $\delta$, corresponding to motions normally associated with the logarithmic region and beyond (attached eddies and superstructures).  (ISA and OSA were originally referred to as small-eddy actuation and large-eddy actuation, respectively, in \cite{marusic2021nature}.)  The lower-frequencies employed for OSA, as compared to ISA, means that OSA can result in positive net power savings.  Furthermore, in contrast to ISA,  the performance of OSA improves with increasing Reynolds number.  \cite{marusic2021nature} attributed this trend to the difference in the turbulent drag composition at high Reynolds numbers, where the relative contribution of large-scale (low frequency) eddies to the wall shear stress increases. 

\subsection{Wall shear stress vs. Reynolds number}

\begin{figure}
	\begin{center}
		\includegraphics[width=1\linewidth,trim = 0mm 0mm 0mm 0mm, clip]{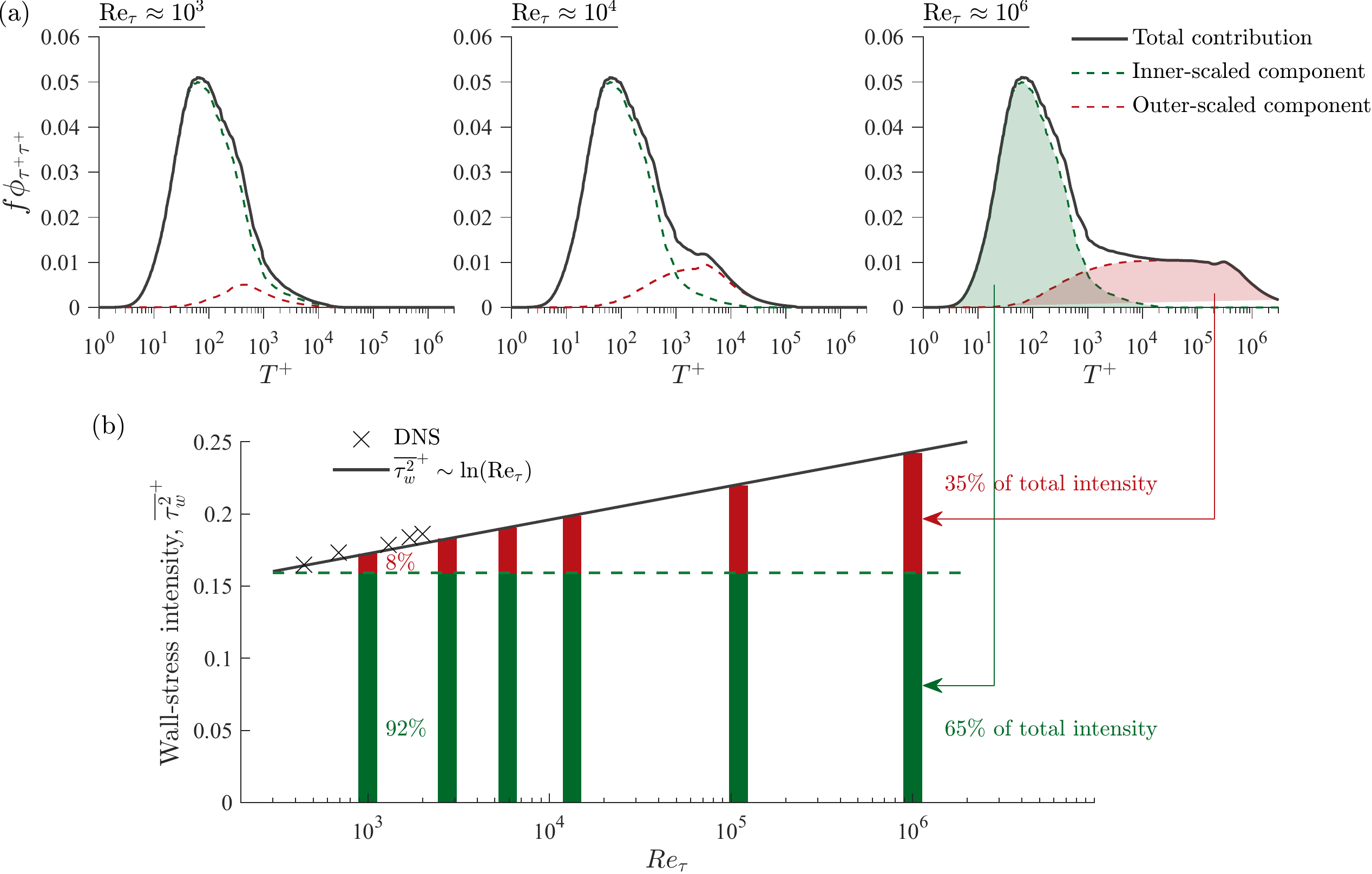}
		\caption{(a) The contributions of near-wall, inner-scaled motions (green) and larger, outer-scaled motions (red) to the pre-multiplied spectra of the wall stress $\tau_w$, computed using predictive models \citep{marusic2010predictive,mathis2013estimating,Chandran2020model}, at $Re_\tau = 10^3,\, 10^4$ and $10^6$. (b) Intensity of wall-shear stress fluctuations as a function of Reynolds number, highlighting the relative contributions from the small-scale and large-scale structures. $\times$, data from direct numerical simulations (DNS) by \citep{jimenez2010turbulent,Sillero2013}.}
		\label{fig:Tau_Model}
	\end{center}
\end{figure}

Figure \ref{fig:Tau_Model}(a) shows the pre-multiplied power spectral density (spectrum) of the wall shear stress, $f \phi_{\tau^+ \tau^+}$, obtained using the predictive models of \cite{marusic2010predictive}, \cite{mathis2013estimating} and \cite{Chandran2020model} at $Re_\tau$ ranging from $10^3$ to $10^6$. Here, $f$ is the frequency and $f^+ = 1/T^+=f \nu /{u^2_{\tau_0}}$.  The spectra show the relative contributions to the wall shear stress from turbulent structures of different time-scales ($T^+$). 
\dc{\cite{mathis2013estimating} used a cut-off frequency of $f^+ = 2.65 \times 10^{-3}\, (T^+ \approx 350)$ to decompose the total wall shear stress spectrum into (i) a Reynolds number invariant, universal contribution from the small, inner-scaled motions ($T^+ < 350$) and (ii) a large-scale contribution from the outer-scaled motions that increased with Reynolds number. Therefore, based on this cut-off time scale of $T^+=350$, we highlight in figure \ref{fig:Tau_Model}(a) the inner-scaled component (shown in green) and the outer-scaled component (shown in red) of the wall shear stress spectra. While the inner-scaled component is the contribution to the wall-stress by the near-wall cycle, the outer-scaled spectra are the contributions from the motions centered in the logarithmic region and above. The latter is obtained here from the phenomenological model of \citet{Chandran2020model} which include hierarchies of self-similar `wall-attached' eddies and VLSMs/superstructures \citep{Kim1999,Hutchins2007a}.}
\dc{As reported by \cite{mathis2013estimating},} \ref{fig:Tau_Model}(a) shows that while the contributions by the inner-scaled motions to the wall shear stress spectra are Reynolds number invariant, the contributions by the inertial outer-scaled motions ($T^+ \gtrsim 350$) increase with increasing Reynolds number. This large-scale trend reflects the growing influence of the energetic outer-region structures on the near-wall turbulence with Reynolds number \citep{marusic2010predictive,Hutchins2007b,agostini2018impact}.  As shown in figure \ref{fig:Tau_Model}(b), the relative contribution of large-scales to the \dc{intensity of wall-shear stress fluctuations, $\overline{\tau_w^2}$,} increases nominally as $\mathrm{ln}(Re_\tau)$, from 8\% at $Re_\tau \approx 10^3$ to about 35\% at $Re_\tau \approx 10^6$. Therefore, at the high Reynolds numbers considered in the present study ($Re_\tau \sim 10^4)$, the outer-scaled contribution is significant, at $18-20\%$ of the total $\overline{\tau_w^2}$.

\subsection{Present study and outline}\label{sec:outline}

In this paper (Part 2) we present experiments to examine streamwise travelling waves of spanwise oscillations as a potential high Reynolds number flow control strategy. We focus only on streamwise travelling waves of oscillations that move in the upstream direction as they have been shown to yield consistent drag reduction  compared to  downstream travelling waves at low Reynolds numbers \citep{quadrio2004critical}.
The combination of the high Reynolds number boundary layer wind tunnel facility at the University of Melbourne and a custom-made surface actuation test bed \citep{marusic2021nature} allows us to study the actuation for a range of parameters ($A^+,\,T_{osc}^+,\,\kappa_x^+$) in the ISA ($T_{osc}^+ \lesssim 350$) and OSA ($T_{osc}^+ \gtrsim 350$) regimes, over the range of friction Reynolds numbers $4500 \le Re_\tau \le 15000$. 

We use multiple experimental techniques (\S \ref{sec:Exp}), including hot-wire anemometry, a drag balance and stereoscopic particle image velocimetry (PIV) to (i) measure changes in skin-friction drag due to the wall actuation and investigate their energy-efficiency, under both ISA and OSA pathways (\S \ref{sec:results:DR_NPS} to \ref{sec:results:local}), and (ii) examine how the wall-actuation affects turbulence statistics and the scale-specific turbulence for a range of wall-heights (\S \ref{sec:Results:meanVel} to \ref{sec:Results:energySpectra}). We specifically focus on the modification of turbulence in the logarithmic region of the boundary layer as this is the major contributor to the bulk turbulence production at high Reynolds numbers \citep{Marusic2010highRe,Smits2011a}.

\section{Experimental techniques}
\label{sec:Exp}
The experiments were conducted in zero pressure gradient boundary layers in the high Reynolds number boundary layer wind tunnel facility \citep{marusic2015evolution} at the University of Melbourne. The wind tunnel has a working section of $27$ m length and a cross-section of $1.89$ m $\times$ $0.92$ m \dc{(width $\times$ height)}. All experiments were conducted at a streamwise location of $x \approx 21$ m, where the boundary layer attains a thickness of $\delta \approx 0.38$ m. 
\dc{Here, $\delta$ is computed by fitting the mean velocity profile to the composite profile of \cite{Chauhan2009}. For the current data set, we found $\delta$ to be approximately $1.26 \times \delta_{99}$, where $\delta_{99}$ is the wall-normal location where the mean streamwise velocity is 99\% of the freestream velocity. (We however note the ambiguity in accurately measuring $\delta_{99}$ and therefore its associated uncertainty \citep{pirozzoli2023outer}.)} 
By varying the freestream velocities between 5 m/s $\leq U_\infty \leq$ 20 m/s, friction Reynolds numbers in the range 4500 $\leq Re_\tau \leq$ 15000 were achieved (see table \ref{tab:exp_param}).

\newcommand{\filsqr}[1]{\tikz{\filldraw[draw=#1,fill=#1] (0,0) rectangle (0.2cm,0.2cm);}}
\definecolor{R100}{rgb}{0.9472, 0.2510, 0.1822}
\definecolor{R10}{rgb}{0.4888, 0.0232, 0.0638}
\definecolor{B50}{rgb}{0.0777, 0.3844, 0.672}
\definecolor{Pu10}{rgb}{0.2687, 0.0415, 0.5108}
\definecolor{Pu70}{rgb}{0.4292, 0.3476, 0.6546}
\definecolor{BR1}{rgb}{0.4668, 0.2665, 0.0334}
\definecolor{BR2}{rgb}{0.7490, 0.5059, 0.1765}
\definecolor{Pu1}{rgb}{0.3965, 0.0148, 0.4583}

\setlength{\tabcolsep}{3.0pt}
\hspace{-1cm}
\begin{table}
	\begin{centering}
	\begin{tabular}{lcccc|cccccc|r}
	Method  & $\mathrm{Re}_\tau$ & \dc{$\mathrm{Re}_\theta$} &$U_\infty$ & $\delta$ & $f$ & $A$ & $\kappa_x$ & $T_{osc}^+$ & $A^+$ & $\kappa_x^+$ & $DR\, (\%)$\\
		&  &  & (m/s) & (m) & (Hz) &(m/s) & ($1/$m)  & &  & &\\
		\hline
		HW, PIV     &  $4500$ &\dc{11750}& 5 & 0.39& $5$ & 0.57 & 20.94 & $405$ & $3.3$ & $0.0018$  &  9 \% \\  
		HW          &  $4500$ &\dc{11750}& 5 & 0.39& $10$ & 1.13 & 20.94 & $203$ & $6.5$ & $0.0018$  &  16 \%\\  
		HW, PIV     &  $4500$ &\dc{11750}& 5 & 0.39& $15$ & 1.7 & 20.94 & $135$ & $9.8$ & $0.0018$  &  19 \%\\  
		HW          &  $4500$ &\dc{11750}& 5 & 0.39& $20$ & 2.26 & 20.94 & $100$ & $13$ & $0.0018$  &  22.3 \%\\  
		\vspace{2 mm}
		HW          &  $4500$ &\dc{11750}& 5 & 0.39& $25$ & 2.83 & 20.94 & $81$ & $16.3$ & $0.0018$  &  21.7 \%\\  
		HW, PIV     &  $6000$ &\dc{15700}& 7 & 0.39& $5$ & 0.57 & 20.94 & $700$ & $2.5$ & $0.0014$  &  6 \%\\  
		HW          &  $6000$ &\dc{15700}& 7 & 0.39& $10$ & 1.13 & 20.94 & $348$ & $4.9$ & $0.0014$  & 10 \% \\  
		HW, DB, PIV   &  $6000$ &\dc{15700}& 7 & 0.39& $15$ & 1.7 & 20.94 & $232$ & $7.4$ & $0.0014$  & 16 \% \\  
		HW          &  $6000$ &\dc{15700}& 7 & 0.39& $20$ &2.26& 20.94 & $174$ & $9.8$ & $0.0014$  & 20 \% \\  
		\vspace{2 mm}
		HW, DB, PIV  &  $6000$ &\dc{15700}& 7 & 0.39& $25$ &2.83& 20.94 & $140$ & $12.3$ & $0.0014$  &  24 \%\\  
		HW &$9700$ &\dc{25200}& 11 &0.39 & $5$ & 0.57 & 20.94 & $1812$ & $1.5$ & $0.0008$ & 5.2 \%\\ 
		HW &$9700$ &\dc{25200}& 11 &0.39 & $10$ &1.13& 20.94 & $906$ & $3$ & $0.0008$ & 9.5 \%\\ 
		HW, DB &$9700$ &\dc{25200}& 11 &0.39 & $15$ &1.7& 20.94 & $604$ & $4.6$ & $0.0008$ & 11.5 \%\\ 
		HW &$9700$ &\dc{25200}& 11 &0.39 & $20$ & 2.26 & 20.94 & $453$ & $6.2$ & $0.0008$ & 12.5 \%\\ 
		 \vspace{2 mm}
		HW, DB &$9700$ &\dc{25200}& 11 & 0.39 & $25$ &2.83& 20.94 & $362$ & $7.8$ & $0.0008$ & 15 \%\\ 
        DB  &$12800$ &\dc{32500}& 15 & 0.385& $15$ &1.7& 20.94 & $1100$ & $3.5$ & $0.0006$ & 8.3 \%\\
        \vspace{2 mm}
		DB &$12800$ &\dc{32500}&15 &0.385& $25$ &2.83& 20.94 & $653$ & $5.7$ & $0.0006$ & 13.3 \%\\ 
		DB &$15000$ &\dc{38000}&20 &0.37& $15$ &1.7& 20.94 & $1975$ & $2.6$ & $0.00046$ & 4.7 \%\\

	\end{tabular}
	\end{centering}
\caption{Summary of experimental parameters. Details of the flow conditions in experiments along with the actuation parameters adopted in the study. The experimental techniques include hot-wire anemometry(HW), drag balance (DB) and stereoscopic particle image velocimetry (PIV). \dc{Here, $U_\infty$ is the freestream velocity and $Re_\theta = \theta U_\infty/\nu$ is the Reynolds number based on momentum thickness ($\theta$). $Re_\tau$ and $Re_\theta$ values mentioned here are for the reference non-actuated conditions. }}
\label{tab:exp_param}
	
\end{table}

\subsection{Surface actuation test bed}
\begin{figure}
\centering
\subfigure{
		\includegraphics[width=1\linewidth,trim = 11mm 68mm 49mm 65mm, clip]{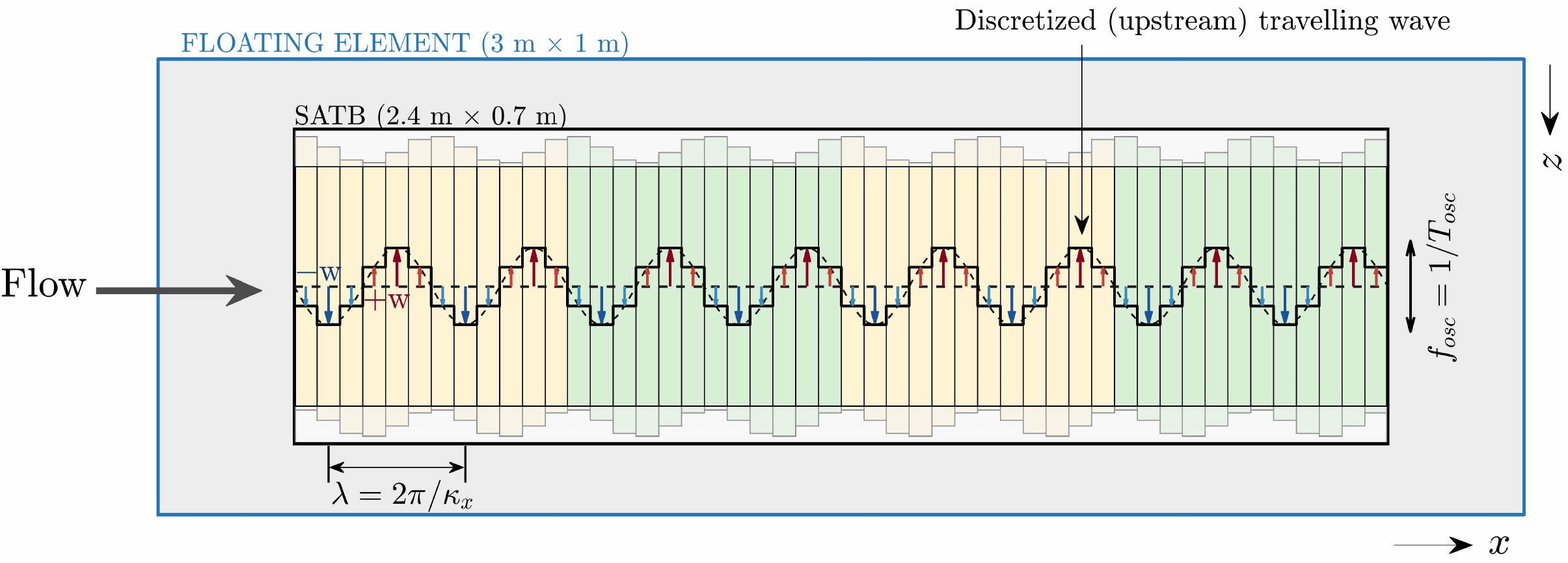}}
  \caption{Schematic of the surface actuation test bed (SATB) installed in the Melbourne wind tunnel facility. The SATB comprises four independently controllable machines (highlighted by different colours) whose synchronous operation generates a discrete facsimile of a long streamwise travelling wave with a total fetch of $8 \lambda$ (= 2.4 m). }
		\label{fig:LFAT_pic}
	
\end{figure}

The streamwise travelling waves of spanwise velocity, as defined by \eqref{wallmotion}, were implemented in the experiments using a surface actuation test bed (SATB). The SATB was custom designed for high-Reynolds number turbulent boundary layers to actuate in both the ISA and OSA regimes, and it measures 2.4 m x 0.7 m, as illustrated in figure \ref{fig:LFAT_pic}. It comprises a series of forty-eight, $50$ mm-wide slats that oscillate in the spanwise direction in a synchronous manner to produce a streamwise travelling, sinusoidal wave that underlies the turbulent boundary layer. Spanwise oscillations within a frequency range 5 Hz $\leq f_{osc} \leq$ 25 Hz were achieved with peak amplitudes (velocities) of actuation equivalent to $A=2 \pi f_{osc}\, d$, where $d= 18$ mm being the fixed half-stroke length. As shown in figure \ref{fig:LFAT_pic}, the streamwise travelling wave is discretized such that six slats constitute a fixed streamwise wavelength with $\lambda = 0.3$ m. \mkf{This level of discretization, though necessary from a practical standpoint, creates a more complex boundary condition with broader spectral content than Eq. \ref{wallmotion} in both the upstream and downstream directions \citep{auteri2010experimental}. At this level of discretization, the mean absolute error of the surface velocities from an ideal sinusoid is approximately $0.17A$, with the edges of the slats experiencing the largest differences, up to a maximum deviation of $0.51 A$. Further, while this discretization introduces higher wavenumber $\mathrm{sinc}$ harmonics of low amplitudes \citep{auteri2010experimental}, we expect their contribution to be limited for the actuation range considered in the experiments. This is validated by a good agreement between the experimental (discrete waveform) and LES (continuous waveform) data at matched conditions, as discussed in \S \ref{sec:results}. }

As highlighted by different colours in the schematic, the SATB was driven by four independently controllable machines, and their selective, synchronous operation enabled a variable streamwise fetch of actuation ($l_{act}$) of $2\lambda \le l_{act} \le 8\lambda$, at $2\lambda$ increments. \dc{Photographs of one of the four independently controllable machines and that of the phase-synchronised assembly of four machines inside the wind tunnel are provided in Appendix \ref{sec:appendix2}}.
\cite{marusic2021nature} provides further details regarding the precision and tolerance associated with the SATB fabrication, together with a supplementary video of its operation in the wind tunnel. 

\subsection{Floating element drag balance}

The SATB is flush mounted on a large-scale floating element drag balance assembly (see figure \ref{fig:LFAT_pic}) that is installed in the bottom wall of the wind tunnel, between 19.5 and 22.5 m downstream of the trip. 
The floating element has an exposed surface area of $3$ m $\times$ $1$ m \citep{baars2016wall}. 
\dc{Due to the long streamwise development length of the boundary layer ($x \approx 21$ m), the boundary layer thickness $\delta$, and therefore $Re_\tau$, varies only marginally (within 4\%; \citealp{talluru2013manipulating}) along the 3 m length of the drag balance for all Reynolds numbers considered here. Therefore, for all cases considered here, we think it is reasonable to treat the boundary layer to be fully developed and its properties (including the skin-friction coefficient) to not vary spatially or temporally along the length of the actuator.}

Compressed air at 10 bar is supplied through four air bearing assemblies to enable a nominally frictionless streamwise movement while any spanwise movements are arrested by additional air bearings. 
A single-beam load cell with a full-scale range of 6 N ($0.06\%$ accuracy of full-scale) constrains the streamwise displacement of the floating element and thereby measures directly the area-averaged skin-friction drag force ($F_w$) on the total exposed surface of the floating element. In-situ calibrations of the load cell with known weights were performed immediately before and after the drag measurements, and the calibrations were performed separately for the actuated and non-actuated cases to account for any preload induced by the mechanical operation of the actuator. 

\begin{figure}
		\begin{center}
			\includegraphics[width=1\linewidth,trim = 00mm 0mm 00mm 00mm, clip]{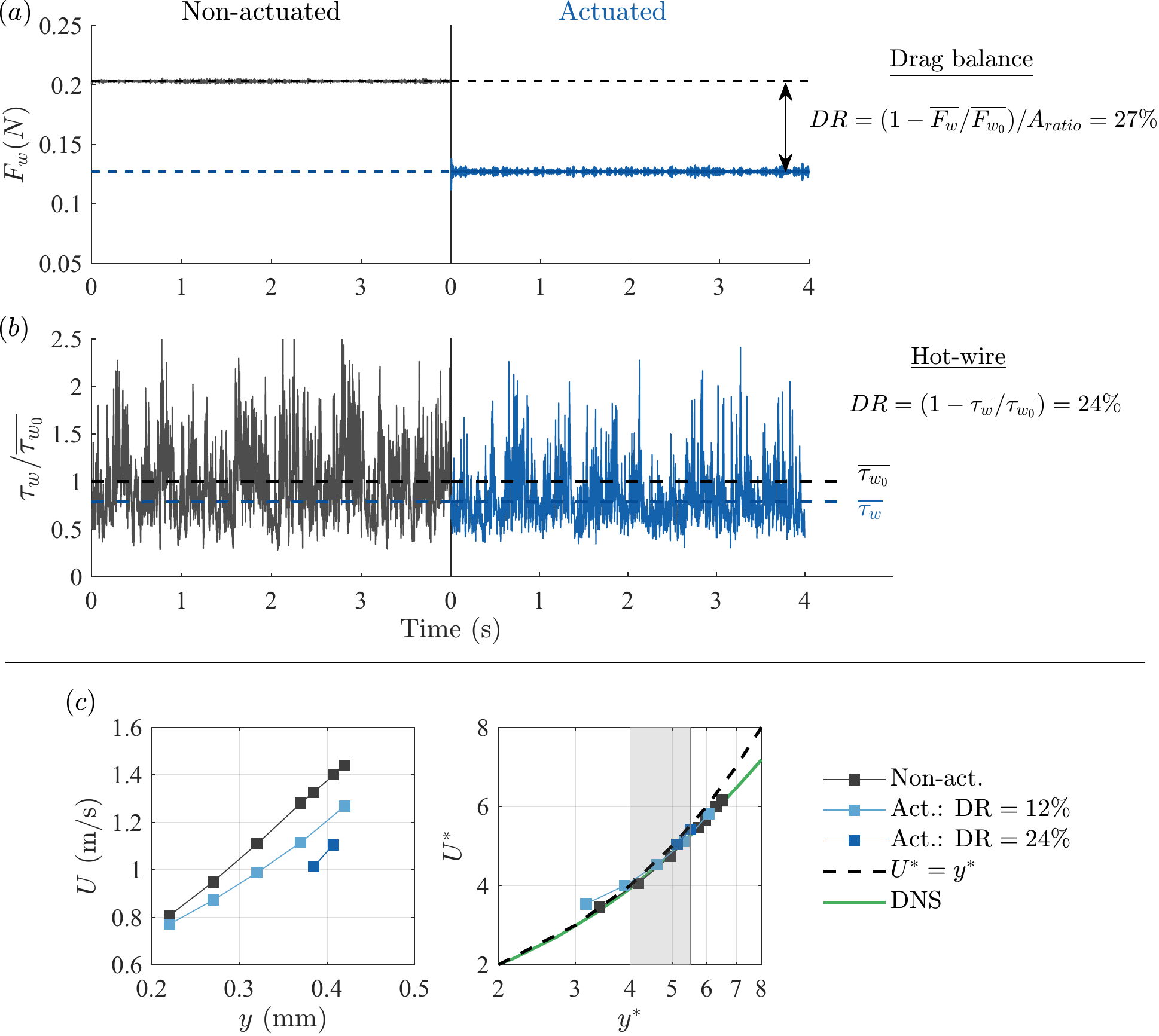}
		\caption{(a) Sample time series of drag measured by the load cell in the drag balance at $Re_\tau = 6000$ for the non-actuated and an actuated case ($A^+ = 12$, $T_{osc}^+=140$, $\kappa_x^+ = 0.0014$) as shown in solid black and blue lines, respectively. The dashed lines denote the time-averaged mean of the respective time signals. (b) Sample time series measurements of wall shear stress $\tau_w$ obtained using hot-wires, for the cases given in (a). (c) Sample mean velocity distributions for the non-actuated and two actuated cases at $Re_\tau = 6000$ obtained using hot wires. Left panel shows profiles of the dimensional mean velocity $U$. Right panel shows the same profiles non-dimensionalized using the actual friction velocity ($U^*=U/u_\tau$).  The grey-shaded region highlights the ``useful linear region'' for $DR$ measurements over SATB. 
		}
		\label{fig:timeseries}
	\end{center}
\end{figure}

The effective drag reduction is therefore estimated according to
\begin{equation}
    DR  = (1-\overline{F_w}/\overline{F_{w_0}})/A_{ratio},
\end{equation}
where $\overline{F_{w_0}}$ and $\overline{F_w}$ are the time-averaged drag forces measured by the load cell for the non-actuated and actuated cases, respectively. The ratio of the actuated area to the total area of the floating element is given by $A_{ratio} = 0.53$, which includes about $8 \%$ of the floating element surface located immediately downstream of the SATB where latent drag reduction effects were present (see figure \ref{fig:LFAT_pic}). 
\dc{However, any latent drag modification on the floating element surface on either sides of the SATB (along the spanwise direction) is neglected and we expect this to be not significant enough since the area-averaged $DR$ measurements from the drag balance are found to be consistent with  both hot-wire and LES data (see section \ref{sec:results}).}
A sample time series of $F_{w_0}$ (black) and $F_{w}$ (blue) is shown in figure \ref{fig:timeseries}(a), for a case with $DR = 27\%$. In all measurements, the signal from the load cell was sampled for at least 60 seconds at 1000 Hz and the drag measurements were repeated at least three times for both the non-actuated and actuated conditions. About 98\% of the repeated drag measurements were within $\pm 0.01$ N of the mean, which resulted in a maximum uncertainty of $\pm 4\%$ in $DR (\%)$, as indicated by error bars in the later plots.

\subsection{Hot-wire anemometry}

Local wall-shear stress measurements were obtained using hot-wire anemometry for $4500 \le Re_\tau \le 9700$. The hot-wire sensors had a diameter and length of $d_{HW}= 2.5$ \textmu m and $l_{HW}=0.5$ mm, respectively, so that $l_{HW}/d_{HW} = 200$ and $5.7 \le l_{HW}^+ \le 12$ \citep{hutchins2009hot, smits_2022} across the $Re_\tau$ range. The sensors were operated using a Melbourne University Constant Temperature Anemometer (MUCTA) at an overheat ratio of 1.8 and a resultant frequency response of 20 kHz. The hot-wire probes were calibrated in-situ in the wind tunnel freestream at 15 different velocities and subsequently fitted with a third-order polynomial. 

The hot-wire measurements were performed in the nominally linear velocity region ($U^* = y^*$) within the viscous sublayer. The instantaneous wall shear stress is proportional to the instantaneous velocity gradient at the wall, and within the linear region this velocity gradient can be approximated to within a few percent error by measuring the streamwise velocity at a single wall-height \citep{Hutchins2002}.  A sample time trace of this deduced wall stress is shown in figure \ref{fig:timeseries}(b). In addition, for measurements at a constant height above the stationary and actuated walls,
\begin{equation}
DR = 1 - \overline{\tau_w}/\overline{\tau_{w_0}} = 1 - U/U_{0}. 
\end{equation}

Figure \ref{fig:timeseries}(c) shows the sample mean velocity from hot-wire data at $Re_\tau = 6000$. In agreement with the observations of \cite{Hutchins2002}, a ``useful linear region'' was found to exist only for a narrow range of wall-heights close to the wall. As highlighted in \ref{fig:timeseries}(c), this region was identified as $4 \le y^* \le 5.5$ over SATB. There, the hot-wire is close enough to the wall to be in the linear region while not close enough for the signal to be contaminated by wall-conduction effects which were observed to come into play for $y^* < 4$. Consequently, in all cases, hot-wire data were acquired at two to four wall-normal locations within the useful linear region that corresponded to about $y \approx 400$ \textmu m at $Re_\tau = 4500$ and $y \approx 200$ \textmu m at $Re_\tau = 9700$. The accuracy of positioning the hot-wires at such close proximity to the wall was determined by the linear optical encoder (RENISHAW RGH24-type, $\pm0.5$ \textmu m accuracy) within the stepper motor-driven vertical traverse, supplemented by a depth measuring displacement microscope (Titan Tool Supply,  $\pm1$ \textmu m accuracy).  The practical challenges associated with identifying the useful linear region restricted the hot-wire measurements to $Re_\tau \le 9700$.

The signals were low-pass filtered using an 8-pole Butterworth filter (Frequency Devices, Inc.\ model 9002) with the roll-off frequency set at half the sampling frequency to minimize aliasing. The signals were sampled at $40$ kHz for $Re_\tau = 4500$ and 6000 and $50$ kHz for $Re_\tau = 9700$. To ensure converged statistics, the signals were sampled for $t=60-90$ seconds, corresponding to non-dimensional boundary layer turnover times of $t U_\infty/\delta = 1100-2500$.

\mkf{Errors in measuring the mean streamwise velocity due to extra cooling from spanwise fluctuations in the Stokes layer can be estimated as 
\begin{equation}
    \epsilon (y) \approx \frac{k^2}{2} \frac{\left<w_{stokes}'^{2}\right>}{ U^{2}}
\end{equation}
\citep{Bruun1995}, where $k$ ($\approx 0.2$) is the hot-wire yaw coefficient and $\left<w_{stokes}'^{2}\right>$ is the spanwise velocity variance due to the Stokes layer. For the range of actuation frequencies and Reynolds numbers encountered in this study, this bias is largest at $Re_\tau = 4500$ and $A^+=16.3$ where the spanwise velocity variance from the Stokes layer is largest and reaches a maximum of $\epsilon < 1.5\%$ at $y^+ = 5$ and $\epsilon < 3.4\%$ at $y^+ = 4$. These errors are even smaller for the other experimental sets of parameters and decrease for smaller values of $A^+$.}

\subsection{Stereoscopic particle image velocimetry (PIV)}
\begin{figure}
	\centering
	\includegraphics[width=0.9\textwidth]{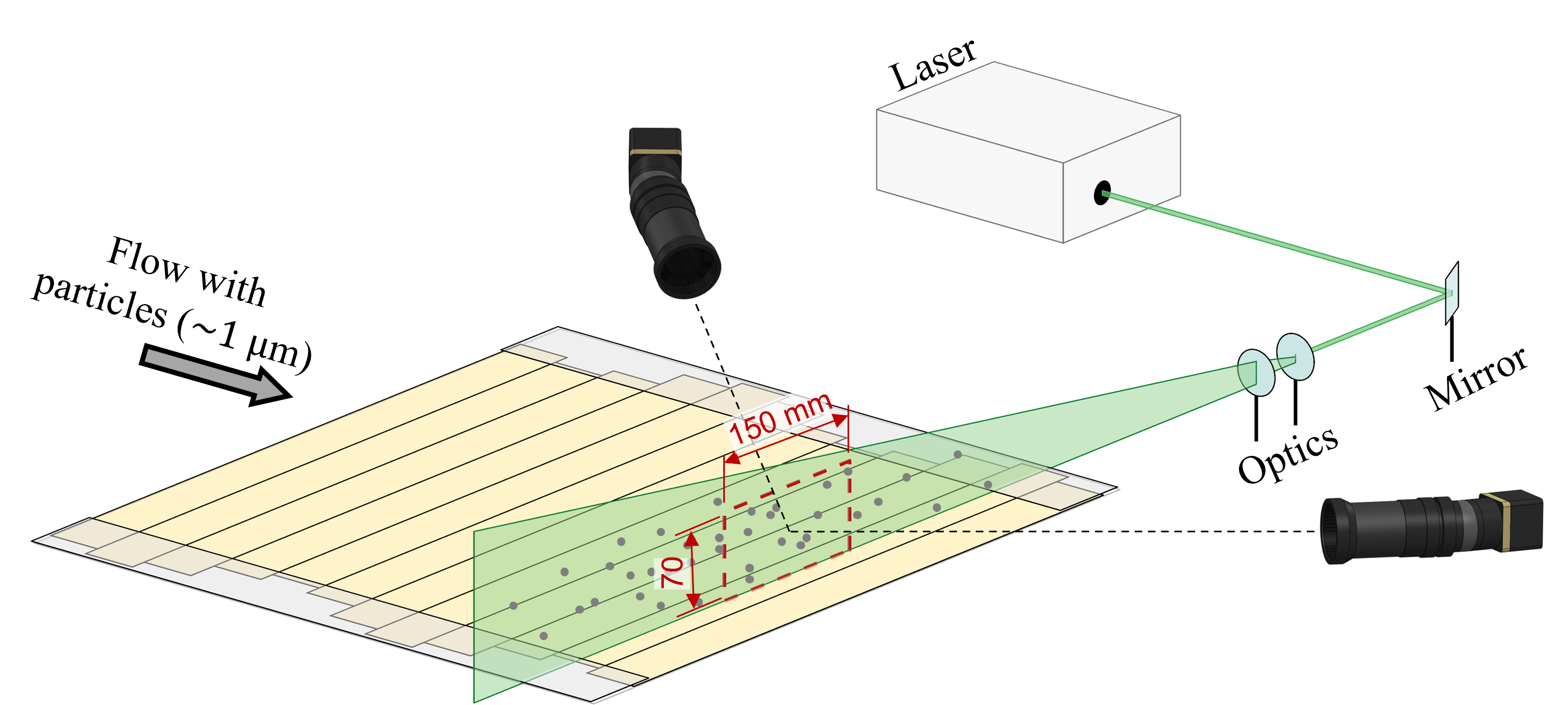}
	\caption{A schematic of the two-camera stereoscopiv-PIV arrangement for measurements over the SATB. The red dashed lines show the field of view of the arrangement ($150$ mm $\times$ $70$ mm) along the spanwise$-$wall-normal plane. }
	\label{fig:SPIV_schematic}
\end{figure}

A two-camera stereoscopic PIV system was used to measure the three velocity components within a 150\,mm\,${\times}$\,70\,mm spanwise ($y{-}z$) plane located about the centre of the tunnel (see figure \ref{fig:SPIV_schematic}). Image pairs were recorded at a constant frequency of 0.5\,Hz using two Imperx GEV-B6620 CCD cameras (6600${\times}$4400 pixels, 5.5 \textmu m pixel pitch, 12 bits per pixel) equipped with Tamron AF 180\,mm lenses at f/8 aperture and Sigma APD 2${\times}$ teleconverters. The cameras were rotated by ${\pm}42^{\circ}$ with respect to the laser sheet and Scheimpflug adapters were used to achieve uniform focus across the measurement plane. A series of cylindrical lenses was used to shape the output of a Spectra Physics Quanta Ray double-pulse Nd:Yag laser (532\,nm, 400\,mJ/pulse) into a 1.5\,mm-thick sheet. Tracer particles of 1-2 \textmu m diameter were generated from a propylene glycol mixture using a fog machine and the tracers were injected upstream of the facility's flow conditioning section for a homogeneous seeding density at the test section.

The acquired images were `dewarped' onto a common grid and two-component velocity fields were computed independently for each camera through cross correlation. We used an iterative deformation method (e.g. \citet{Astarita2005}) with Blackman weighted interrogation regions of size 48${\times}$48 pixels and an overlap of 75\%, resulting in a vector grid spacing of 12 pixels (${\approx}$0.25\,mm). The three-component velocity fields were finally computed by combining the velocity fields from each camera using the local viewing angles obtained through the calibration procedure. The camera configuration provided two estimates of the vertical velocity component, which were used to reduce the noise contribution to the vertical velocity variance $\langle {v^\prime}^2 \rangle$ \citep{Cameron2013}.

The cases covered by the PIV measurements at $Re_\tau = 4500$ and $6000$ are reported in table \ref{tab:exp_param}.
A total of 1000 independent velocity fields were captured for each case. The spatial resolution of the PIV data was assessed by  comparing the streamwise variance distributions obtained for the non-actuated cases with the hot-wire data from \cite{marusic2015evolution} at matched $Re_\tau$ (see figure \ref{fig:PIV_stress}). 
Although PIV experiments were also carried out at $Re_\tau = 9700$, the data were found to have insufficient spatial resolution for the current investigation and are therefore not included in the paper.

\section{Results and discussion}
\label{sec:results}

\subsection{Drag reduction and net power savings}
\label{sec:results:DR_NPS}
\mkf{For a statistically-stationary, streamwise homogeneous flow, the averaged wall shear stress depends on the following parameters:}
\begin{equation}
	\frac{\overline{\tau_w}}{\rho} = u_\tau^2 = g_1(U_\infty,\, \delta,\, \nu,\, \kappa_x,\, T_{osc}, A). 
	\label{eq:tau_g1}
\end{equation}
Dimensional analysis thereby yields
\begin{equation}
	DR = g_2(\kappa_x^+,\, T_{\mathrm{osc}}^+,\, A^+,\, Re_\tau).
	\label{eq:Dim_DR}
\end{equation}
(This is the same as equation 1.3 in Part 1.)
For the SATB test-bed, $A^+ = (2\pi/T_{\mathrm{osc}}^+)(d/\delta) Re_\tau$ and $\kappa_x^+ = (2\pi/Re_\tau)(\delta/\lambda)$, where $d/\delta$ and $\delta/\lambda$ are approximately constant as $\delta = 0.38 \pm 0.01$ m across all measurements. 
Therefore, $A^+  \propto \, Re_\tau/T_{osc}^+$ and $\kappa_x^+ \, \propto \, 1/Re_\tau$.
Consequently, $Re_\tau$ could not be varied independent of the parameters of actuation ($A^+,\,T_{osc}^+,\,\kappa_x^+$). Similarly, at a particular $Re_\tau$, one actuation parameter could not be varied independently of the others.  Therefore, in our experiments, based on the above relationship between $A^+,\,T_{osc}^+$ and $Re_\tau$, \eqref{eq:Dim_DR}  simplifies to
\begin{subequations}
   \begin{align}
    DR &= g_3(T_{\mathrm{osc}}^+,\, Re_\tau), \, \mathrm{or} \label{eq:DR_vs_A_T_a}\\
    DR & ~\approx~ g_4(A^+). \label{eq:DR_vs_A_T_b}
    \end{align}
\end{subequations}
\begin{figure}
	\centering
	\includegraphics[trim = 0mm 5mm 0mm 0mm, clip, width=1\linewidth]{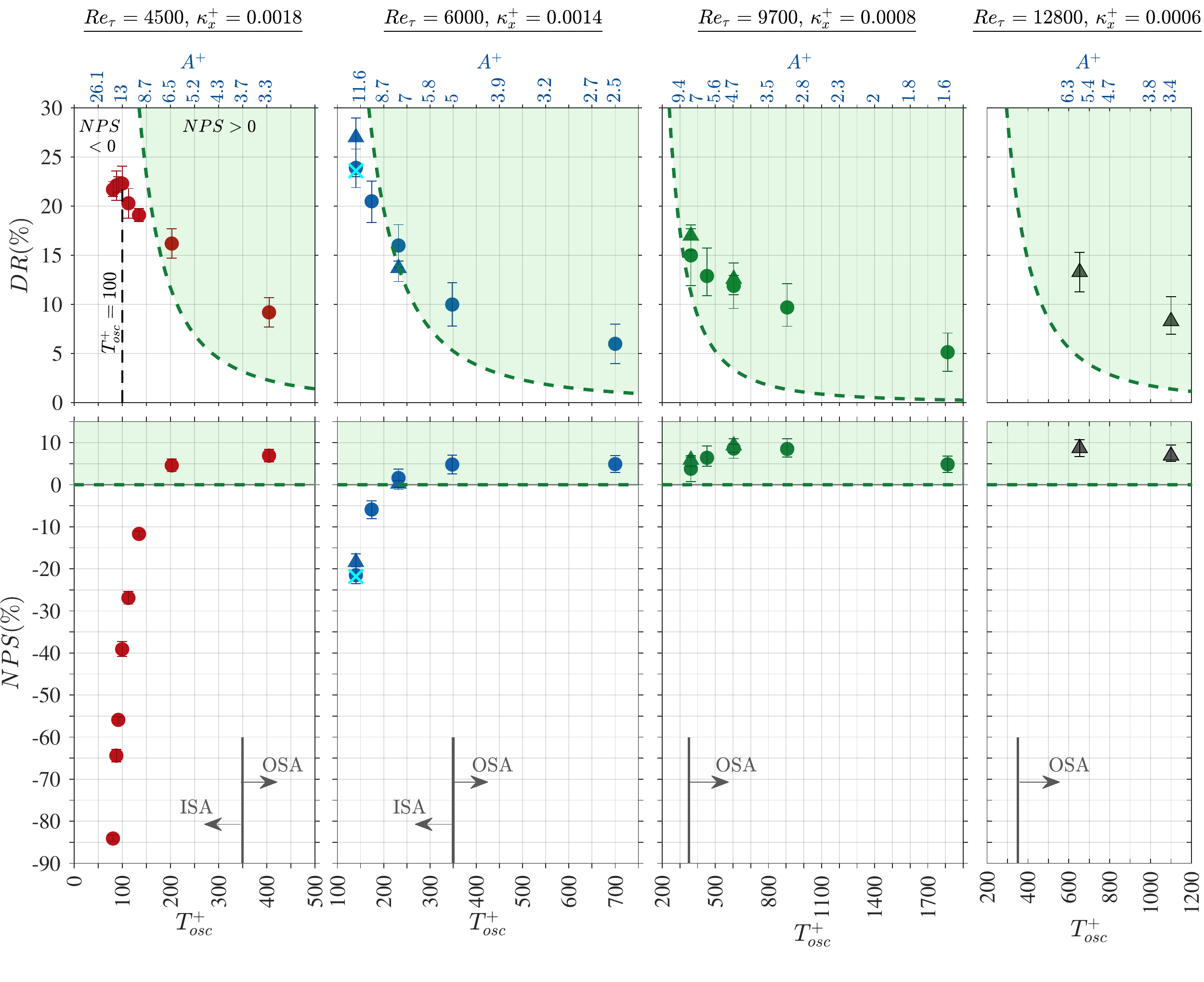}
	\caption{Drag reduction ($DR$) and net power savings ($NPS$) as functions of $T_{osc}^+$ and $A^+$ for $Re_\tau$ ranging from 4500 to 12800. The green shaded regions correspond to $NPS > 0$. The circles and triangles represent the hot-wire data and drag balance data, respectively. The cross indicates an LES data point at $Re_\tau = 6000$ at matched actuation conditions. The error bars indicate one standard deviation uncertainty ranges.}
	\label{fig:DRvsAT}
\end{figure}

Figure \ref{fig:DRvsAT} shows how the measured drag reduction ($DR$) and net power savings ($NPS$) depend on $T_{osc}^+$ and $A^+$. Here, $NPS$ is computed using the generalised Stokes layer (GSL) theory \citep{quadrio2011laminar,gatti2013performance}, $NPS$ being the difference between $DR$ and  the net power required to move the flow sideways, as in \eqref{wallmotion}, to generate the Stokes layer. \mkf{In other words, it is the net input power required by an `ideal' actuation system to implement \eqref{wallmotion}, i.e., neglecting any mechanical losses}. 
\dc{Refer to \S\ 3.6 in Part 1 for further details of how net input power is computed using the GSL theory and its validation with respect to LES data.
Specifically, figure 10(b) in Part 1 shows that for the low-wavenumber, low-frequency (long-time period) actuation range, GSL over-predicts the input power for actuation when compared to LES. A similar observation  was also made by \cite{touber2012near} for long-time period actuation. Therefore, we note that GSL could plausibly serve as an underestimate for the idealized net power savings, especially for the OSA cases.}

The results demonstrate that at $Re_\tau = 4500$ a peak $DR \approx 24\%$ is achieved with ISA at $T_{osc}^+ \approx 100$ and $A^+ \approx 13$. The time scale of this actuation corresponds to the time scale of the near-wall streaks that contribute to the peak in figure \ref{fig:Tau_Model}. However, by factoring in the power input for the actuation, this ISA case actually incurs negative net power savings ($NPS$) of $-40\%$ (i.e., a net power cost). We see that for most cases in the ISA regime, NPS is negative, with the loss increasing sharply for $T_{osc}^+ <100$. In contrast, when the time period of actuation is increased beyond $T_{osc}^+ \gtrsim 350$ to target the OSA pathway, positive net power savings of $ +7\%$ are achieved with a moderate $DR = 9\%$. A similar trend is observed as $Re_\tau$ is increased to 6000 and 9700, where OSA consistently results in positive net power savings in the range $4\% \le NPS \le 9\%$ for corresponding drag reductions in the range $17\% \ge DR \ge 5\%$. It is important to note that the positive $NPS$ are achieved here with relatively modest amplitudes, $1.5 \le A^+ \le 7.8$.  
\dc{Further, here we note that for matched actuation parameters (see table \ref{tab:exp_param}), the $DR$ measured through hot-wire (circles) and drag balance (triangles) show good agreement with any difference being within the experimental uncertainty.}
The LES reference point at $Re_\tau = 6000$ (reported in Part 1) for matched actuation parameters also agrees well with the data obtained from hot-wire and drag-balance measurements.

\dc{At higher Reynolds number, the energy-efficiency offered by the OSA pathway improves further. 
This was discussed in detail by \citet{marusic2021nature} who showed, from LES and experiments at fixed dimensionless actuation parameters, that $DR$ and $NPS$ increase with increasing Reynolds numbers, opposite in trend to low-Reynolds number results for the same matched parameters. This trend for the OSA pathway was attributed to the contribution of large, outer-scaled motions to the wall shear stress, which  while carrying very little energy at low Reynolds numbers ($Re_\tau \sim 10^3$) becomes significant at high Reynolds numbers (refer to figure \ref{fig:Tau_Model}b).} 
For example, at $Re_\tau = 12800$ the OSA pathway results in $NPS = 8.7\%$ and 7\%, corresponding to $DR = 13.3\%$ and 8.3\%, respectively. At the highest Reynolds number achieved here, $Re_\tau = 15000$, a very low-amplitude ($A^+ = 2.6$), low-frequency ($T_{osc}^+ = 1975$), low power actuation yielded $NPS = 4\%$ even with a relatively low amount of drag reduction, $DR = 4.6\%$. 

\begin{figure}
	\centering
	\includegraphics[trim = 7mm 0mm 30mm 0mm, clip, width=1\linewidth]{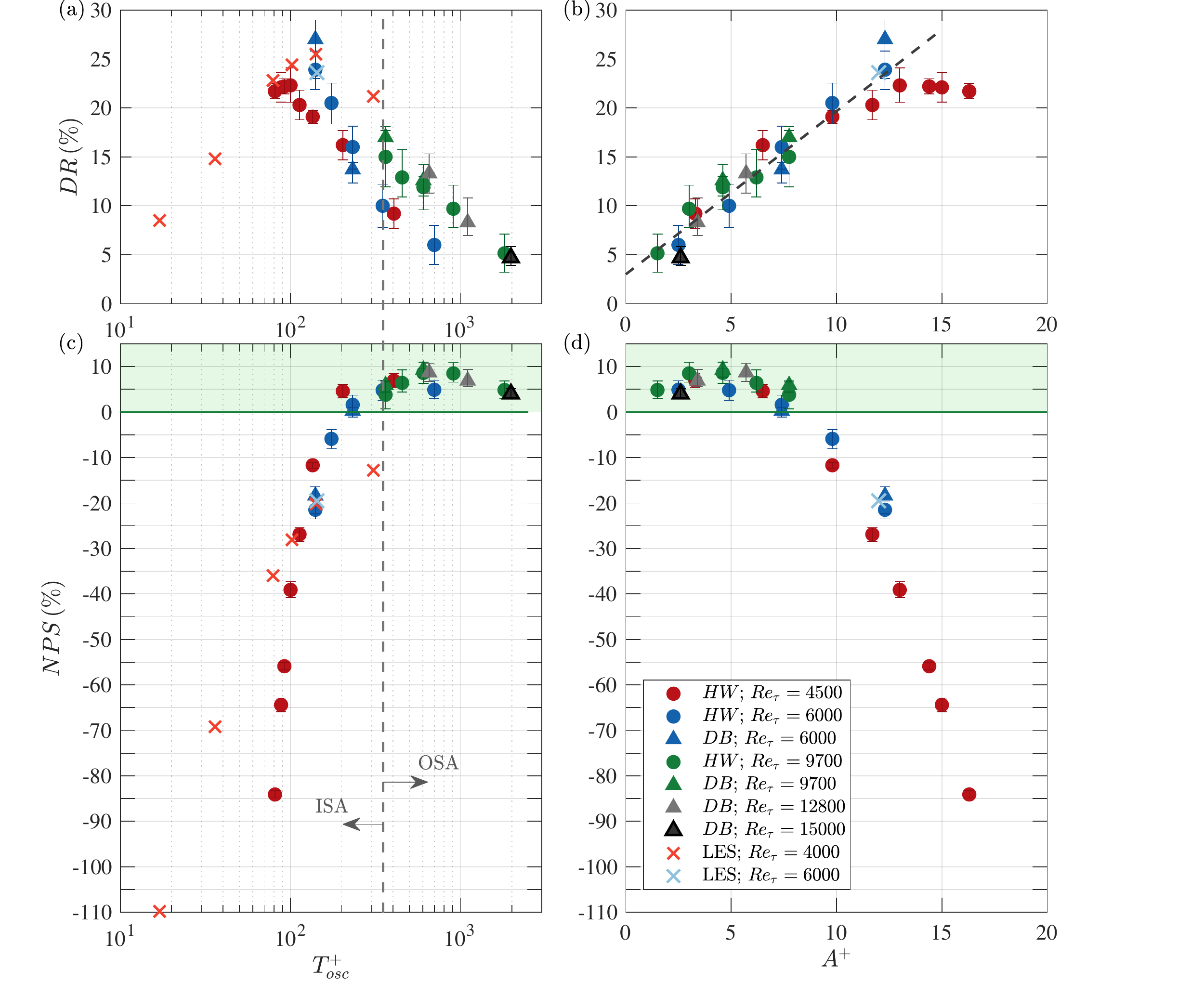}
	\caption{Drag reduction $DR$ and net power savings $NPS$ versus $T_{osc}^+$ and  $A^+$ across the full range of Reynolds numbers and actuation parameters. The LES results are from Part 1.}
	\label{fig:DR_summary}
\end{figure}

The complete $DR$ and $NPS$ results are presented in figure~\ref{fig:DR_summary}, along with select LES data points from Part 1. Figure \ref{fig:DR_summary}(a) shows the distribution of $DR$ as a function of $T_{osc}^+$ and $Re_\tau$, as given by \eqref{eq:DR_vs_A_T_a}. Again, we see that the experimental data agree well with the two LES data points at similar operating conditions, i.e., at $Re_\tau \approx 4500$, $T_{osc}^+ \approx 100$ and at $Re_\tau = 6000$, $T_{osc}^+ = 140$. Our data show that $DR$ varies only moderately with Reynolds number under both ISA and OSA strategies, for the range investigated. However, as noted earlier, in our experiments, as $T^+_{osc}$ increases $A^+$ decreases and therefore limits $A^+$ to relatively smaller values at high Reynolds numbers (see table \ref{tab:exp_param}).

Figure~\ref{fig:DR_summary}(a) shows that the drag reduction peaks at $DR \approx 25\%$ at $T^+_{osc} \approx 100$ in the ISA regime, beyond which $DR$ decreases approximately semi-logarithmically with increasing $T^+_{osc}$.  The $DR$ in LES decreases by a lesser extent than in the experiments because $A^+$ in LES is fixed at $12$, while in the experiments it decreases from about 12 at $T^+_{osc} \simeq 100$ to $2.6$ at $T^+_{osc} = 1975$. However, these high-frequency ISA actuations mostly yield negative $NPS$, as shown in figure \ref{fig:DR_summary}(c), with $NPS$ decreasing rapidly as $T_{osc}^+$ falls below 100. $NPS$ is found to be slightly positive (up to $+4\%$) for two data points at $T_{osc}^+ \approx 200$. In contrast, the OSA pathway consistently yields positive $NPS$ in the range $5 - 10\%$ even with a moderate $DR$ of $5 - 15\%$.

The variation of $DR$ and $NPS$ with $A^+$ is shown in figures~\ref{fig:DR_summary}(b,d). For the current actuation system, $DR$ when expressed as a function of $A^+$ is nearly independent of Reynolds number \eqref{eq:DR_vs_A_T_b}. The results support this conclusion, in that for a particular $A^+$, $DR$ obtained at various $Re_\tau$ collapse well.  Further, $DR$ increases almost linearly with $A^+$ up to $A^+ \approx 12$, beyond which it appears to saturate.  Although the value of $A^+$ to achieve maximum $DR$ appears to be $\ge 12$, the plot of $NPS$ in figure \ref{fig:DR_summary}(d) suggests lower amplitudes of actuation ($A^+ \lessapprox 7$) are necessary to yield positive $NPS$. These lower amplitudes correspond to larger time periods of actuation as $A^+ \propto  Re_\tau/T_{osc}^+$ in our experiments.
We believe these trends demonstrate the potential of an OSA pathway, with its low-frequency, low-amplitude actuation strategy, for energy-efficient drag reduction at high Reynolds numbers. 

\subsection{Spatial modification of turbulent drag and its recovery}
\label{sec:results:local}

\begin{figure}
	\centering
	\subfigure{
	\includegraphics[width=0.9\textwidth]{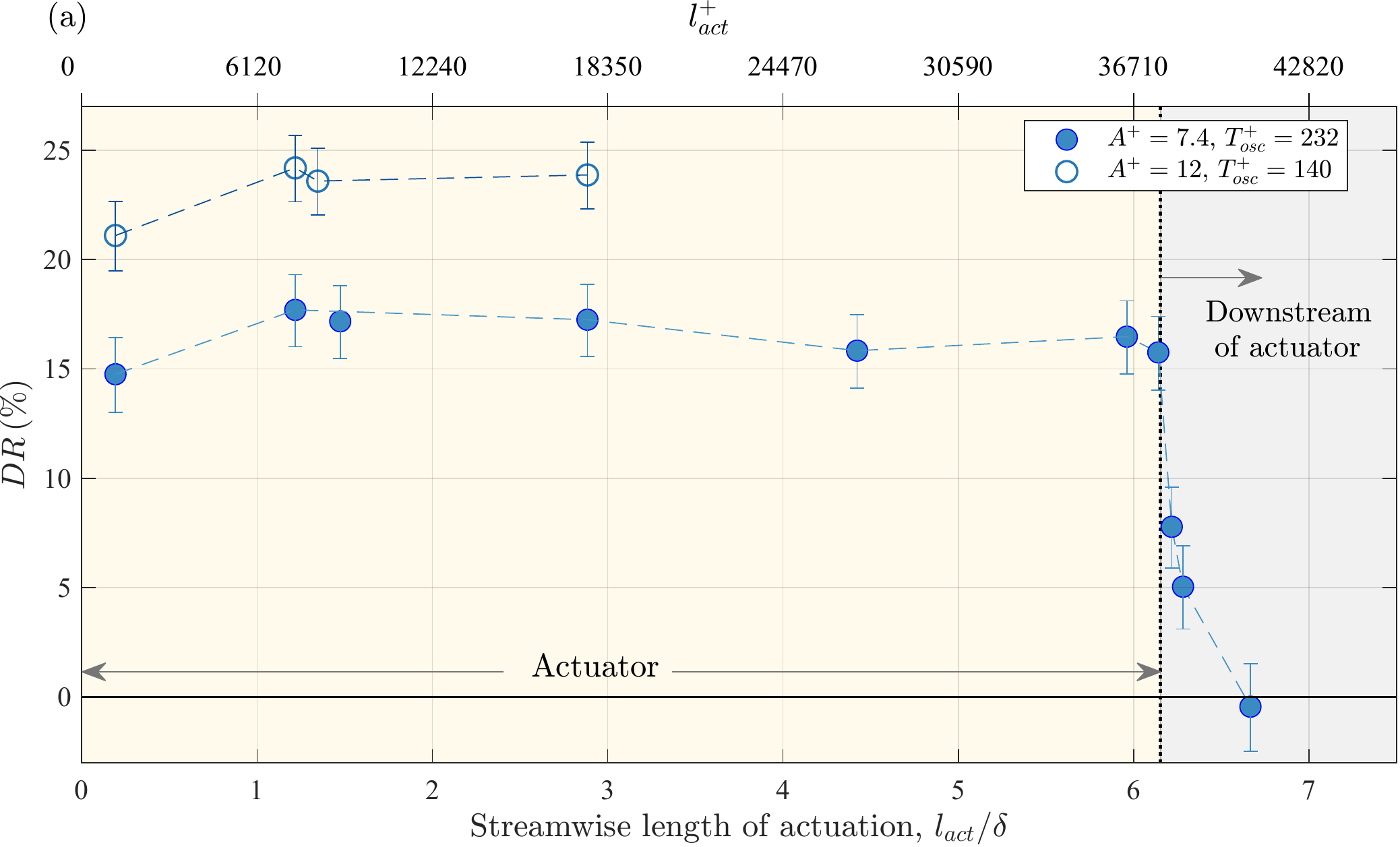}} 
	\subfigure{
	\includegraphics[width=0.9\textwidth]{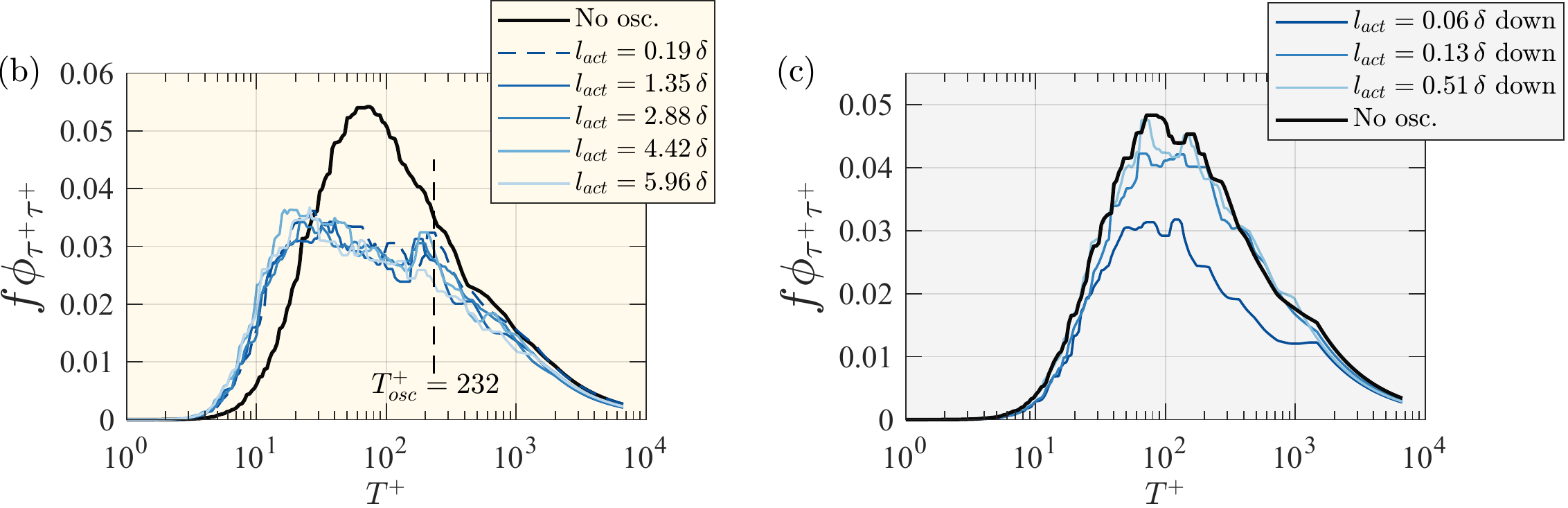}} 
	\caption{(a) Variation of drag reduction ($DR$) along the length of the actuator, $0<l_{act}<2.4$ m (yellow shaded region) at $Re_\tau = 6000$ for two cases: $A^+ = 7.4,\, T_{osc}^+ = 232$ and $A^+ = 12,\, T_{osc}^+ = 140$. For the first case, the recovery of skin friction drag downstream of the actuator $l_{act}>2.4$ m (grey shaded region) is also plotted. (b) The effect of the streamwise length of actuation on the spectra of $\tau_w$ for $l_{act}$ up to 2.4 m ($\approx 6\, \delta$) for the case $A^+ = 7.4,\, T_{osc}^+ = 232$. (c) The downstream recovery of $\tau_w-$spectra to the reference non-actuated state for the case in (b).}  
	\label{fig:DR_vs_actlen}
\end{figure}

The variation of the drag reduction with downstream distance, obtained using hot wire anemometry, are shown figure \ref{fig:DR_vs_actlen}(a) for two ISA cases. The majority (85\%) of the net $DR$ is achieved within $l_{act} \approx 0.1$ m, which corresponds to $l_{act}/\lambda = 0.33$ (or $l_{act}/\delta = 0.25$), where $l_{act}$ is the length of actuation. In viscous units, this is equivalent to $l_{act}^+ \approx 1500$ and therefore 50\% greater than the nominal length of the near-wall streaks. This trend of rapidly rising $DR$ with actuation length agrees with the observations from the temporal wall-oscillation studies of \citet{ricco2004effects} and \citet{skote2019wall}\mkf{, as well as with similar observations from the standing-wave wall-oscillations of \citet{skote2011}}.  However, the localised $DR$ is observed to saturate by $l_{act} = 2\lambda = 0.6$ m, beyond which the drag reduction is almost constant with $DR \approx 16\%$ for $T_{osc}^+=232$ and $DR \approx 24\%$ for $T_{osc}^+=140$.  The wall shear stress spectra also collapse for 0.075 m $\leq l_{act} \leq$ 2.325 m (see figure \ref{fig:DR_vs_actlen}(b)), which suggests that in addition to the time-averaged $DR$, the broadband time-scales of turbulent fluctuations also saturate within $l_{act} \approx 0.1$ m. Therefore, all hot-wire data that were used to compute $DR$ (\S\ref{sec:results:DR_NPS}) were acquired with an actuation length of at least $l_{act} = 2\lambda$.

Similarly, the recovery of skin-friction drag downstream of the actuator is very rapid.  For the $T_{osc}^+=232$ case shown in  figures \ref{fig:DR_vs_actlen}(a) and (c), a complete recovery of both mean drag and the spectra of wall-stress is observed within 0.2 m ($\approx$ 3000 viscous units) downstream of the actuator. This recovery length is similar to that observed by \citet{ricco2004effects} and \citet{skote2019wall} for $T_{osc}^+ = 67$ in the ISA regime. 

The spatial transients could not be investigated for OSA.  It was not possible to achieve a matched $A^+ \, (\approx 10)$ for OSA at similar $Re_\tau$, due to the limitations of the current experimental setup. We consider this to be an important subject for a future study.

\subsection{Mean velocity} \label{sec:Results:meanVel}
\begin{figure}
	\centering
	\includegraphics[trim = 0mm 0mm 0mm 0mm, clip, width=1\linewidth]{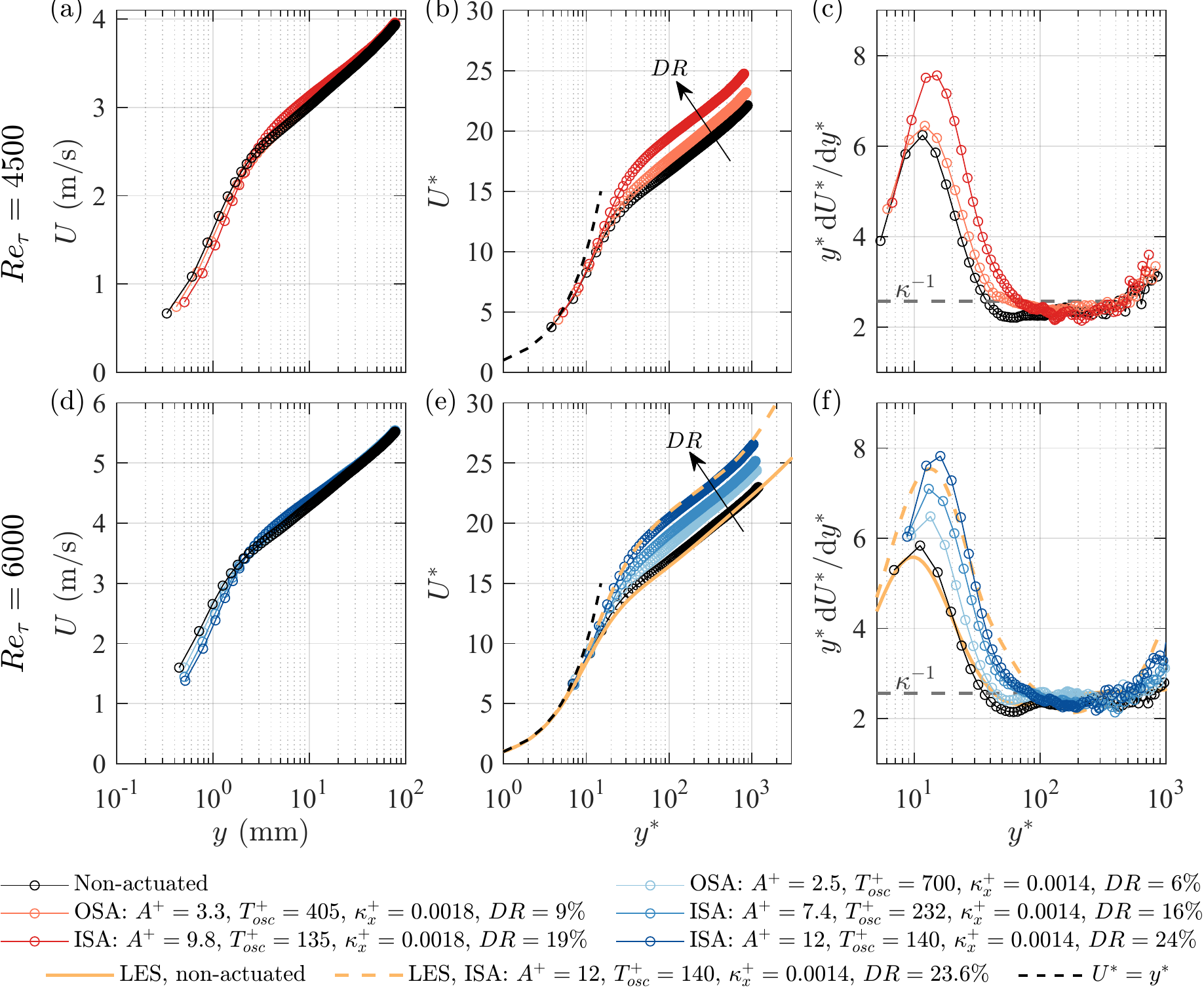}
	\caption{(a,d) Wall-normal profiles of mean streamwise velocity, $U$, for the non-actuated and actuated cases at $Re_\tau = 4500$ and 6000, respectively . (b,e) Profiles in (a,d) normalized using the local $u_\tau$: $U^* = U/u_\tau$ and $y^* = y u_\tau/ \nu$. (c,f) Diagnostic function ($y^*\,\mathrm{d} U^*/\mathrm{d}y^*$) for the non-actuated and actuated cases.  The LES data are from Part 1 at $Re_\tau = 6000$ \citep[Appendix A]{rouhi2022part1}. 
	} 
	\label{fig:PIV_meanVel}
\end{figure}
\dc{The profiles of mean streamwise velocity obtained from the stereo-PIV measurements are shown in figure~\ref{fig:PIV_meanVel}.  Here, the mean velocity profiles are obtained by averaging across the spanwise direction and time. The stereo-PIV data are acquired over the actuated wall and with a streamwise actuation length of $l_{act}=4\lambda$, where the modified skin friction due to the actuation was found to be invariant with downstream distance (see section \ref{sec:results:local}).} 
As displayed in figure~\ref{fig:PIV_meanVel}(a), the dimensional mean velocity reduces systematically near the wall for the drag-reduced cases, consistent with previous studies on streamwise-travelling waves \citep{gatti2016reynolds,bird2018experimental}. 

In figures~\ref{fig:PIV_meanVel}(b,e) the velocity profiles are normalized with $u_\tau$, i.e., $U^* = U/u_\tau$ and $y^* = y u_\tau/ \nu$, \dc{where the local $u_\tau$ is estimated by fitting the PIV sublayer profiles to $U^* = y^*$ (the values matched the hot-wire values to within 3\%).} As a consequence, the normalized velocity profiles are forced to agree near the wall. 
In the outer region, however, the profiles for the drag-reduced cases are shifted upwards systematically, with the shift proportional to the amount of drag reduction. 
These shifted profiles have a nominally constant slope of $\kappa^{-1}$ for all cases considered here. \cite{skote2014scaling} reported that the slope of the mean velocity profile varies as $(\kappa\, \sqrt{1-DR})^{-1}$ in the initial non-equilibrium state and subsequently reverts back to $\kappa^{-1}$ when the whole of the boundary layer has adjusted to the wall actuation and hence reached an equilibrium state.   
This is further verified in figures \ref{fig:PIV_meanVel}(c,f) using a log-law diagnostic function, $y^*\, \mathrm{d}U^*/\mathrm{d}y^*$ \citep{nagib2007approach,skote2019wall}. The profiles of both non-actuated and actuated cases collapse in the logarithmic region ($200 \lesssim y^* \lesssim 0.1Re_\tau$), with a nearly constant $\kappa^{-1}$. Furthermore, the shift in the peak of the diagnostic function to higher $y^*$ indicates the thickening of the viscous sublayer with increasing drag reduction, as reported in Part 1. In figures \ref{fig:PIV_meanVel}(e,f), the reduced-domain LES data at $Re_\tau = 6000$ (Part 1, Appendix A) are included for the non-actuated case and an actuated case at matched conditions ($A^+=12$, $T_{osc}^+=140$) where $DR = 23.6\%$. These profiles agree well with the PIV data.

\subsection{Turbulence statistics}
\label{Sec:TurbStat}

\begin{figure}
	\centering
	\includegraphics[trim = 12mm 0mm 10mm 10mm, clip, width=1\linewidth]{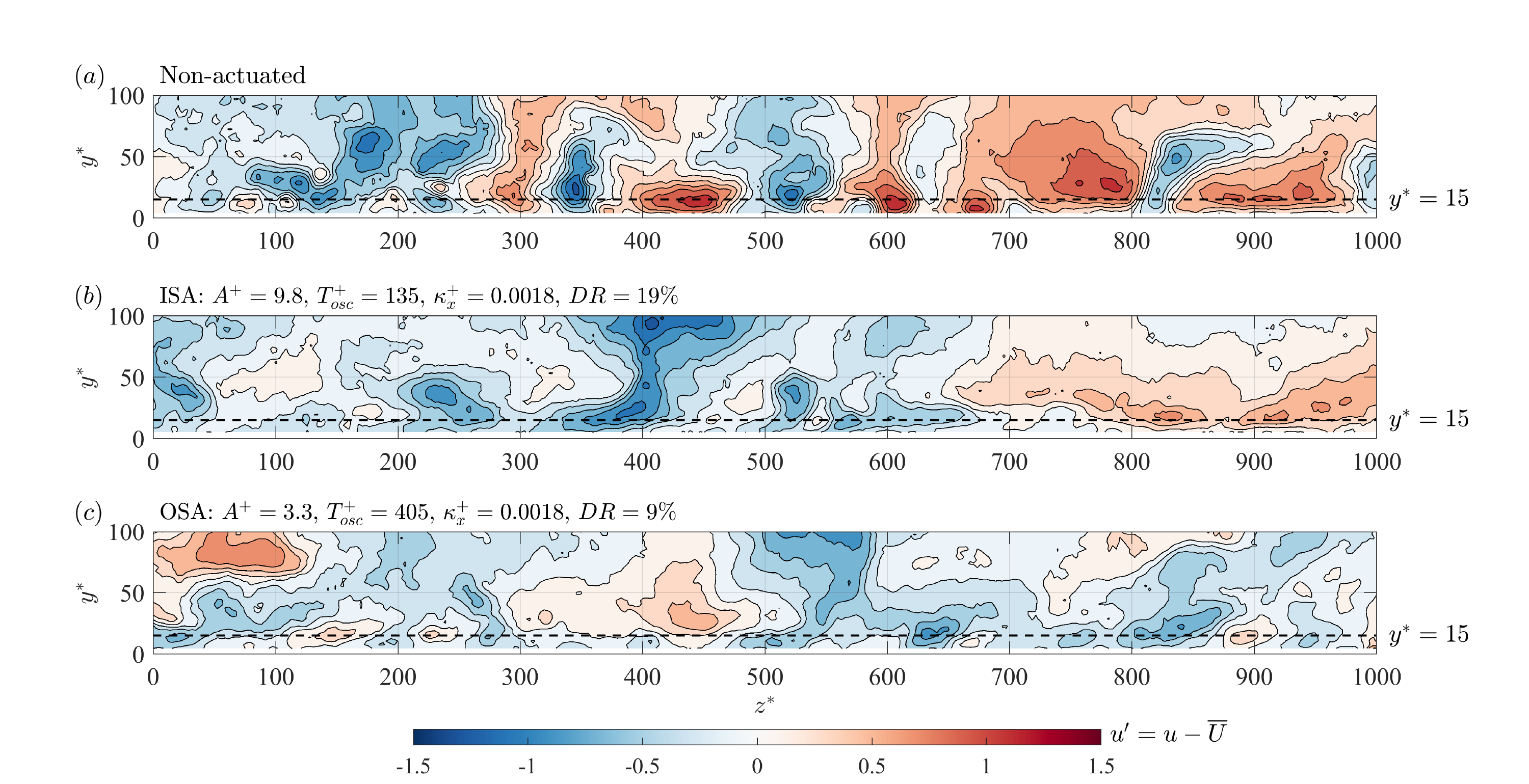}
	\caption{Spanwise$-$wall-normal planes showing the fluctuating streamwise velocity, $u'$, at $Re_\tau = 4500$ for the (a) non-actuated and (b) an ISA case and (c) an OSA case. 
	}  
	\label{fig:PIV_visual}
\end{figure}

Figure \ref{fig:PIV_visual} show sample snapshots of the fluctuating streamwise velocity from the PIV measurements. For the non-actuated case, we see the alternating regions of positive and negative velocity fluctuations (red and blue, respectively) that are characteristic of the near-wall streaks. These well-documented streaks are centered around $y^* = 15$ with a mean spanwise separation of about 100 viscous units, as seen in the figure. For the actuated cases, \dc{both ISA and OSA}, the near-wall streaks are clearly weakened, and might be taken as a symptom of reduced drag. This effect of the actuation on the near-wall turbulence was discussed in Part 1, and with respect to the LES data shown earlier in figure \ref{fig:LES_illus}.
\dc{We note that figure \ref{fig:PIV_visual}(a-c) are only sample snapshots of the flow, provided here to visualize how the actuation affect the instantaneous flow structures near the wall. In order to quantify this effect of actuation on the turbulent fluctuations, we compute the mean turbulence statistics.}

\begin{figure}
	\centering
	\includegraphics[width=1\textwidth]{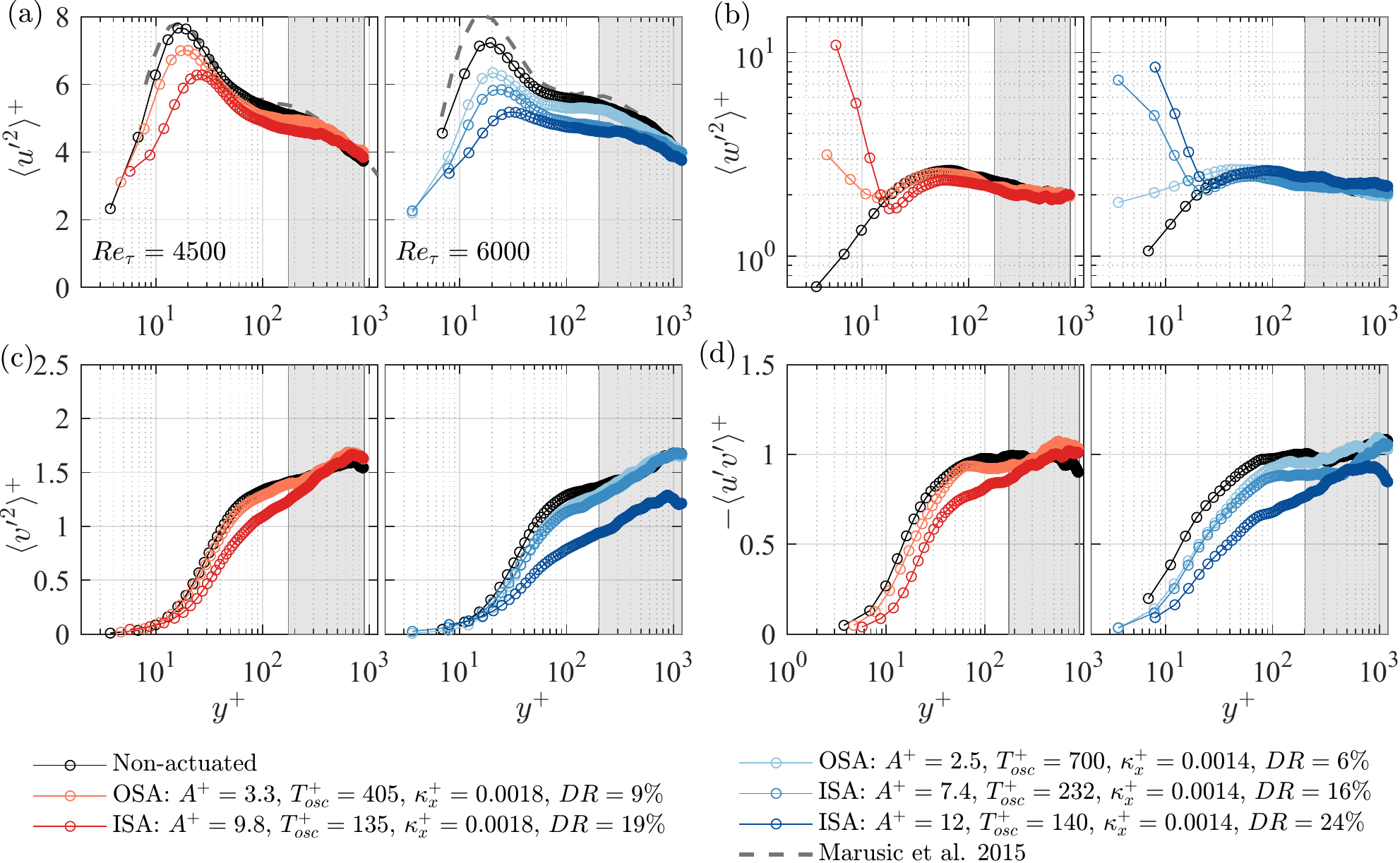}
	\caption{(a-c) Normal stresses and (d) Reynolds shear stress for the non-actuated and actuated cases at $Re_\tau =4500$ and 6000. The axes are normalized using the reference $u_{\tau_0}$. The grey-shaded region refers to the logarithmic region. The profiles of $\langle {w^\prime}^2 \rangle^+$ are plotted in a log-log scale in (b) to highlight the extent of its near-wall amplification due to the Stokes layer (as represented in Part 1). The $\langle {u^\prime}^2 \rangle^+$ profiles from the hot-wire measurements of \cite{marusic2015evolution} at matched $Re_\tau$ are included in (a) as reference for the non-actuated cases and to highlight the spatial resolution of PIV data.}  
	\label{fig:PIV_stress}
\end{figure}

Figure 11 shows the corresponding Reynolds stress statistics from the PIV experiments, averaged across both the spanwise direction and time. These statistics are often scaled either with the reference non-actuated $u_{\tau_0}$ (`$+$' superscript) or with the actual $u_\tau$ (`$*$' superscript). Scaling with $u_{\tau_0}$ highlights the absolute response of the boundary layer to the wall-actuation \citep{laadhari1994turbulence,baron1995turbulent,ricco2004effects,touber2012near,agostini2014spanwise}, while scaling with the actual $u_\tau$ tests the universal nature of these drag-modified turbulence statistics (\citealp{baron1995turbulent,choi2002near,touber2012near,quadrio2011laminar,skote2013,gatti2016reynolds}). \S3.3 in Part 1 presents a detailed discussion on these different scaling choices and their key differences. 
Here, in Part 2, our priority is to understand the extent of the high-Reynolds-number boundary layer that is impacted by the current wall-actuation strategy, in an absolute sense. We find this especially interesting as high-Reynolds-number boundary layers are characterized by large separation of scales and strong inner-outer interactions. 
Therefore, here we choose $u_{\tau_0}$ to normalize the turbulence statistics. However, for reference, a plot of turbulent stresses normalized by the actual $u_\tau$ is provided in Appendix \ref{sec:appendix1}.
 
As shown in figures~\ref{fig:PIV_stress}(a-d), all components of the Reynolds stress are affected by the actuation, with the strongest attenuation observed for the highest drag reducing case. In particular, the peak in $\langle {u^\prime}^2 \rangle^+$ is  increasingly attenuated and shifted to higher $y^+$ as $DR$ increases, in accord with previous studies at lower $Re_\tau$  (\cite{jung1992suppression,choi2001mechanism,ricco2004effects,quadrio2011laminar,touber2012near,skote2013} etc.). Similarly, significant attenuation is observed in the profiles of $\langle {v^\prime}^2 \rangle^+$ and  $-\langle u^\prime v^\prime \rangle^+$, indicating a reduced momentum flux towards the wall \citep{ricco2021review}. For $\langle {w^\prime}^2 \rangle^+$, however, the periodic oscillations of the Stokes layer produce a sharp increase near the wall ($y^+ \lesssim 25$). 
These trends in the Reynolds stresses are consistent with the discussion in Part 1 (\S3.3) where we also subsequently decomposed $\langle {w^\prime}^2 \rangle^+$ into the phase-averaged and the stochastic components, and investigated the interaction between the Stokes layer and the background turbulence. A similar analysis could not be performed with the PIV data as the number of images per phase were inadequate to generate converged phase-averaged statistics.

\begin{figure}
	\centering
	\includegraphics[width=1\textwidth]{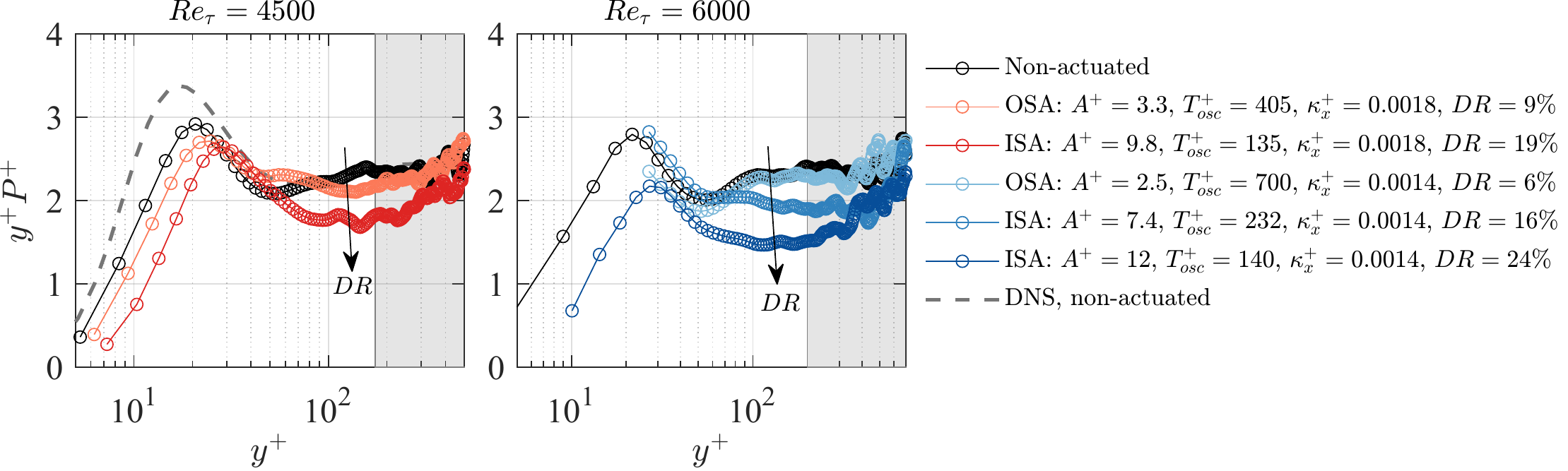}
	\caption{Premultiplied turbulence production for the non-actuated and actuated cases at $Re_\tau =4500$ and 6000. Here, the normalization is using the reference $u_{\tau_0}$ and the grey-shaded region refers to the logarithmic region. The DNS data of \cite{Lozano2014} at a similar Reynolds number is included as a reference for the non-actuated case at $Re_\tau = 4500$. 
	}  
	\label{fig:PIV_prod}
\end{figure}

Figure \ref{fig:PIV_stress} demonstrates that the effects of the actuation persist away from the wall into the logarithmic region (grey shaded region). To highlight these effects, we show in figure \ref{fig:PIV_prod} the pre-multiplied turbulent kinetic energy production $y^+ P^+$, where the production term $P^+ = -\langle u^\prime v^\prime \rangle^+ \, dU^+/dy^+$.  Here, equal areas under the curve (on this log-linear plot) indicate equal contributions to the production. The logarithmic region is known to be an important contributor at high Reynolds numbers \citep{MMH10}. Interestingly, the wall actuation reduces the turbulence production everywhere, near the wall and in the logarithmic region (consistent with figure 5e in Part 1). The attenuation in the logarithmic region shows the efficacy of the current wall actuation strategy to influence the inertial-scale dynamics at high Reynolds numbers.
The reduction of turbulence production in this outer region occurs for both ISA ($T_{osc}^+ \lesssim 350$) and OSA ($T_{osc}^+ \gtrsim 350$) cases and the reduction is stronger with increasing $DR$. 

For our study, the experimental restrictions imposed by the SATB hardware and the PIV technique, limit the $A^+$ values and $Re_\tau$ range that can be examined. In figure \ref{fig:PIV_prod}, the OSA cases have $A^+=2.5$ and 3.3 while ISA cases are at $A^+ \ge 7.4$. Given that $DR$ was observed to increase almost linearly with $A^+$ in the range $A^+ \lesssim 12$ (see figure \ref{fig:DR_summary}), we expect that OSA would be even more effective at larger amplitudes of actuation.

\subsection{Energy spectra} \label{sec:Results:energySpectra}

\begin{figure}
	\centering
	\subfigure{
	\includegraphics[trim = 0mm 20mm 80mm 0mm, clip, width=0.9\linewidth]{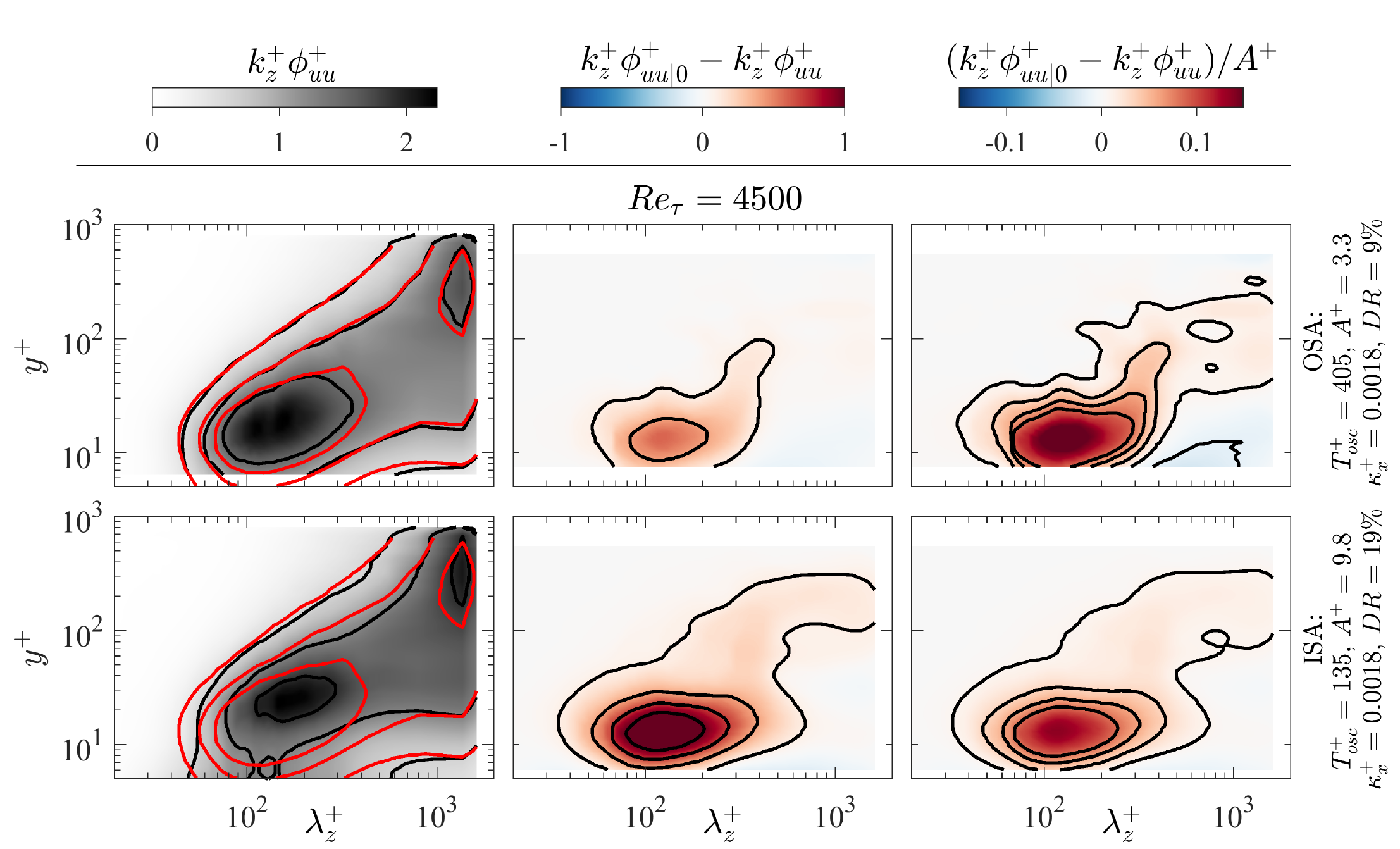}} 
	\subfigure{
	\includegraphics[trim = 0mm 0mm 80mm 0mm, clip, width=0.9\linewidth]{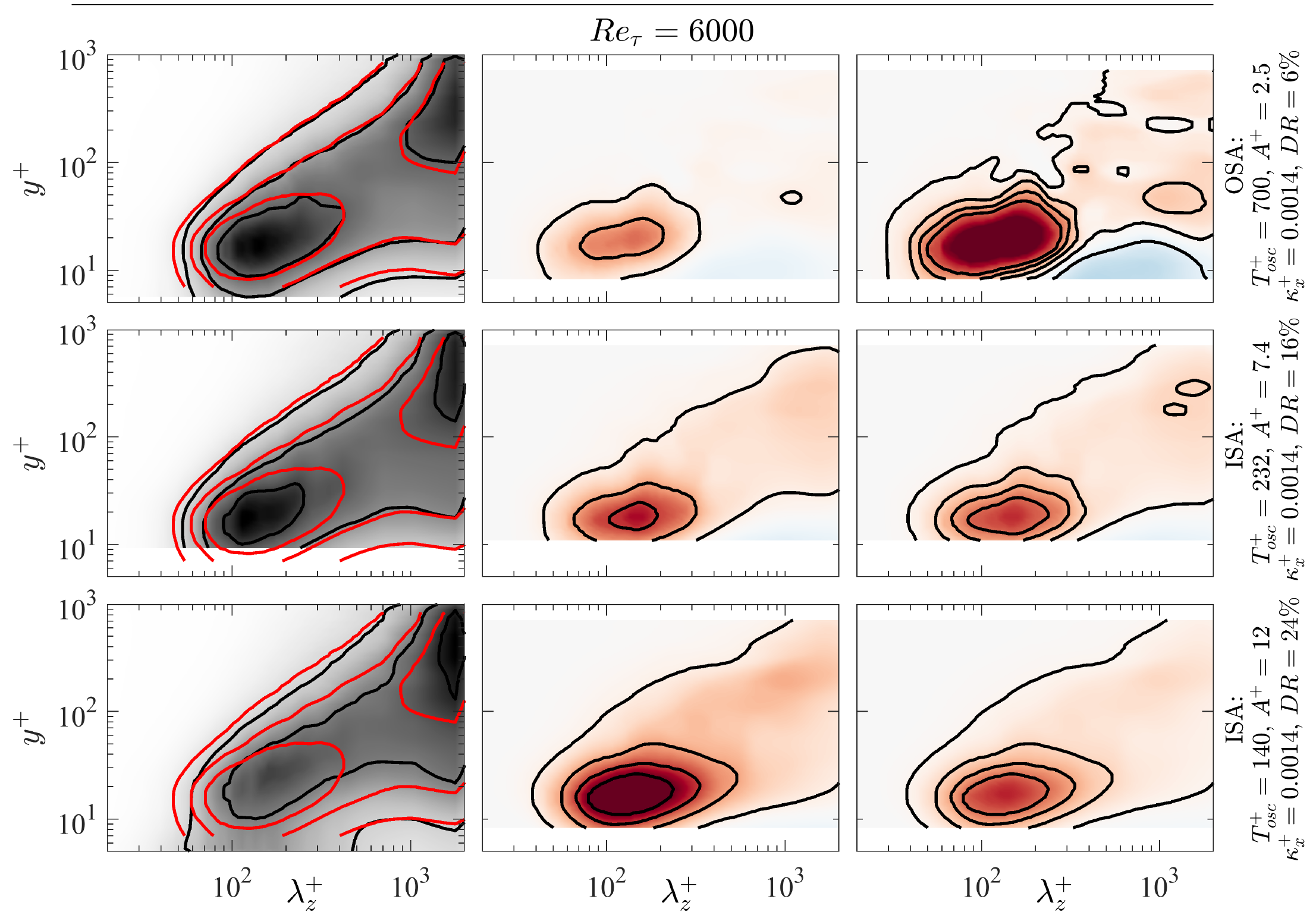}} 
	\caption{Premultiplied spectrograms of streamwise velocity $k_z^+ \phi_{uu}^+$, computed from PIV data, as functions of spanwise wavelength ($\lambda_z$) and wall-height. The left panels show the spectra for the non-actuated (red) and actuated (black) contour conditions. The middle panels show the difference between the non-actuated and actuated spectra (the red-shaded contours  indicate energy attenuation), and the right-most panels show the difference normalized by the respective amplitude of actuation $A^+$. Here, the spectra and the axes are normalized using the reference $u_{\tau_0}$.}  
	\label{fig:PIV_uu_spectra}
\end{figure}

\begin{figure}
	\centering
	\subfigure{
	\includegraphics[trim = 0mm 20mm 80mm 0mm, clip, width=0.9\linewidth]{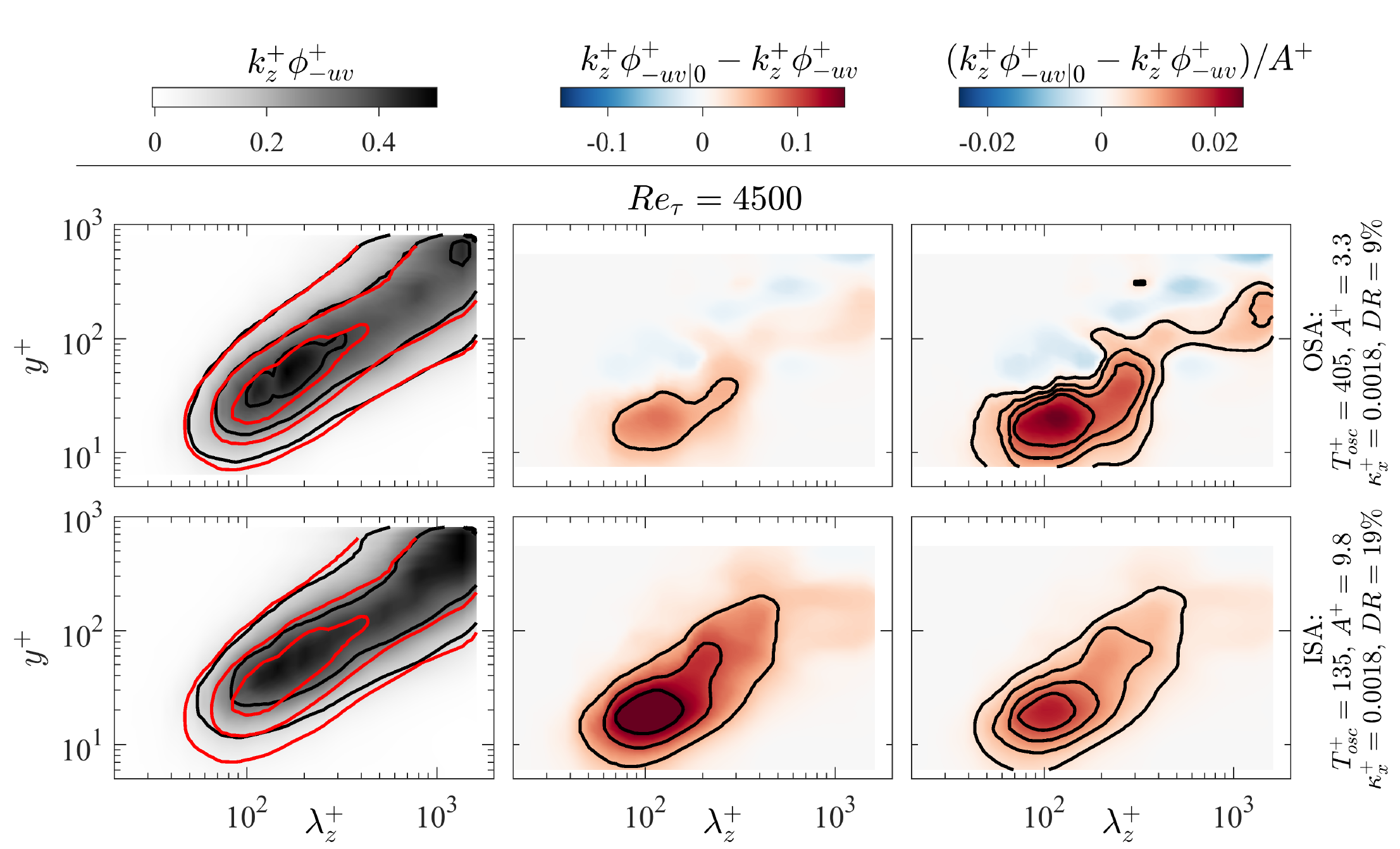}} 
	\subfigure{
	\includegraphics[trim = 0mm 0mm 80mm 0mm, clip, width=0.9\linewidth]{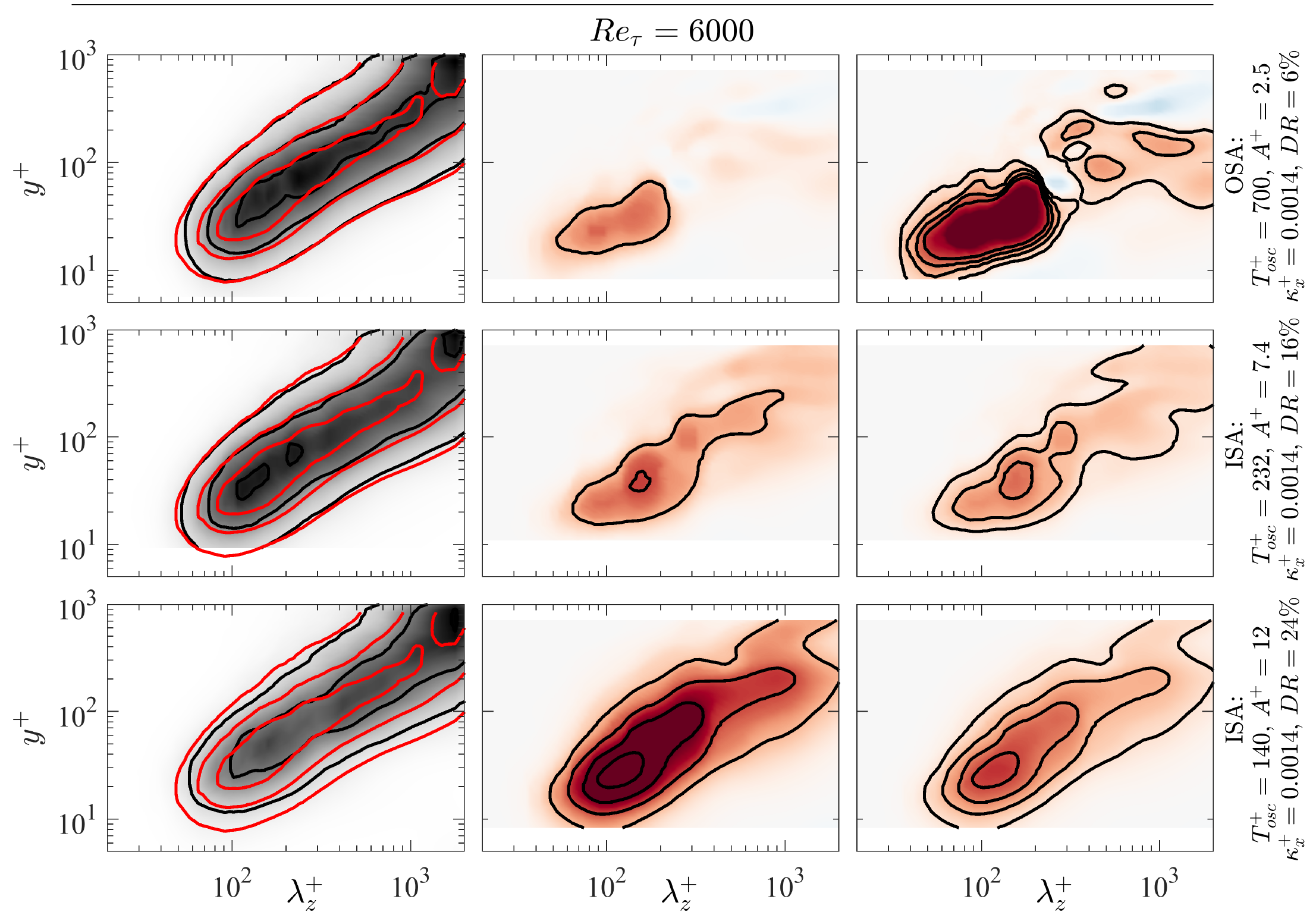}} 
	\caption{Premultiplied co-spectra $k_z^+ \phi_{-uv}^+$, computed with the PIV data, as functions of spanwise wavelength ($\lambda_z$) and wall-height. Here, the spectra and the axes are normalized using the reference $u_{\tau_0}$. The figure is panelized as in figure \ref{fig:PIV_uu_spectra}.}  
	\label{fig:PIV_uv_spectra}
\end{figure}

The spanwise/wall-normal PIV fields were used to compute the spanwise-premultiplied energy spectra of the streamwise velocity component, $k_z^+ \phi_{uu}^+$, and the co-spectra, $k_z^+ \phi_{-uv}^+$ as a function of spanwise wavelength ($\lambda_z$) and wall-height ($y^+$). Here, $k_z^+ \phi_{uu}^+ = k_z \phi_{uu}/u_{\tau_0}^2$ and $k_z^+ \phi_{-uv}^+ = k_z \phi_{-uv}/u_{\tau_0}^2$. The streamwise turbulence intensity $\langle {u^\prime}^2 \rangle^+$ and the Reynolds shear stress $-\langle u^\prime v^\prime \rangle^+$ are the integral of $k_z^+ \phi_{uu}^+$ and $k_z^+ \phi_{-uv}^+$, respectively, as given by,
\begin{equation}
    \langle {u^\prime}^2 \rangle^+=\int_{-\infty}^\infty k_z^+\phi_{uu}^+ \, \text{d}(\mathrm{ln}\,\lambda_z^+) \,\,\,\, \mathrm{and} \,\,\,\, -\langle u^\prime v^\prime \rangle^+= \int_{-\infty}^{\infty} k_z^+\phi_{-uv}^+ \, \text{d}(\mathrm{ln}\,\lambda_z^+).
    \label{eq:spectra}
\end{equation}
The energy spectra allow us to study the scale-specific response to the wall-actuation, revealing the range of scales that contribute to the attenuation of turbulence intensities observed in \S \ref{Sec:TurbStat}.

The spectrograms of $k_z^+ \phi_{uu}^+$ are displayed in the left-hand panels of figure \ref{fig:PIV_uu_spectra}. 
For the non-actuated cases, the spectrograms show the expected near-wall peak at $y^+ \approx 15$ and $\lambda_z^+ \approx 100$ of the near-wall cycle. The spectrograms also show the emergence of a secondary peak in the logarithmic region, typically observed at high Reynolds numbers \citep{Hutchins2007a,Mathis2009,Vallikivi2015_Spectra,lee2015direct,samie2018fully}. 
With actuation, the near-wall peak is attenuated and its location shifts to slightly higher wall-heights, corresponding with the shift in the peak in $\langle {u^\prime}^2 \rangle^+$ seen earlier (figure \ref{fig:PIV_stress}). A similar trend was observed in Part 1 in the spectrograms of streamwise velocity at $Re_\tau = 4000$, and by \cite{gatti2018spectra} at lower Reynolds numbers. 

To highlight changes due to actuation, the spectrograms from the actuated cases are subtracted from the corresponding non-actuated spectrograms and plotted in the middle panels in figure~\ref{fig:PIV_uu_spectra}. The red-shaded contours therefore indicate energy attenuation. We see that the actuation results in the attenuation of a broad range of turbulent scales and importantly, that the effect of actuation at the wall is also felt by the large-scales ($\lambda_z^+ \gg 100$) in the outer region. The extent of this effect appears to be directly proportional to $DR$.  
Since $DR$ was found to vary almost linearly with $A^+$ for $A^+ \le 12$ (figure \ref{fig:DR_summary}), we divide the difference spectra by $A^+$ in the right-most panels in the figure. (Note, for the SATB the values of $A^+$ decrease with increasing $T_{osc}^+$ as $A^+ = (2\pi/T_{osc}^+)(d/\delta)Re_\tau$.) In these scaled plots, the level of attenuation, along with the range of turbulent scales ($\lambda_z$) and wall-heights that are attenuated, look more similar among the different cases at a given Reynolds number. 

Similar trends are observed in the co-spectra, $k_z^+ \phi_{-uv}^+$, as shown in figure \ref{fig:PIV_uv_spectra}. The wall actuation attenuates the energy of the spectral peaks while their wall-normal locations are shifted to slightly higher wall-heights. These trends support the notion that the actuation reduces the momentum transfer to the wall \citep{luchini1996reducing,gatti2018spectra}. As in  figure \ref{fig:PIV_uu_spectra}, when the difference co-spectra are divided by $A^+$ (right-most panels), the level and the range of scales attenuated look more similar across the different cases at a particular $Re_\tau$.
If this conjecture of $A^+$ scaling holds, then at matched amplitudes  ISA and OSA could potentially have a similar impact on the range of turbulent scales attenuated. This would make the low-frequency OSA a good deal more attractive due to its significantly lower input power requirements compared to ISA.  However, testing this conjecture would require varying $A^+$ independent of $T_{osc}^+$, which is not possible with the current actuation system.

\begin{figure}
	\centering
	\includegraphics[width=1\textwidth]{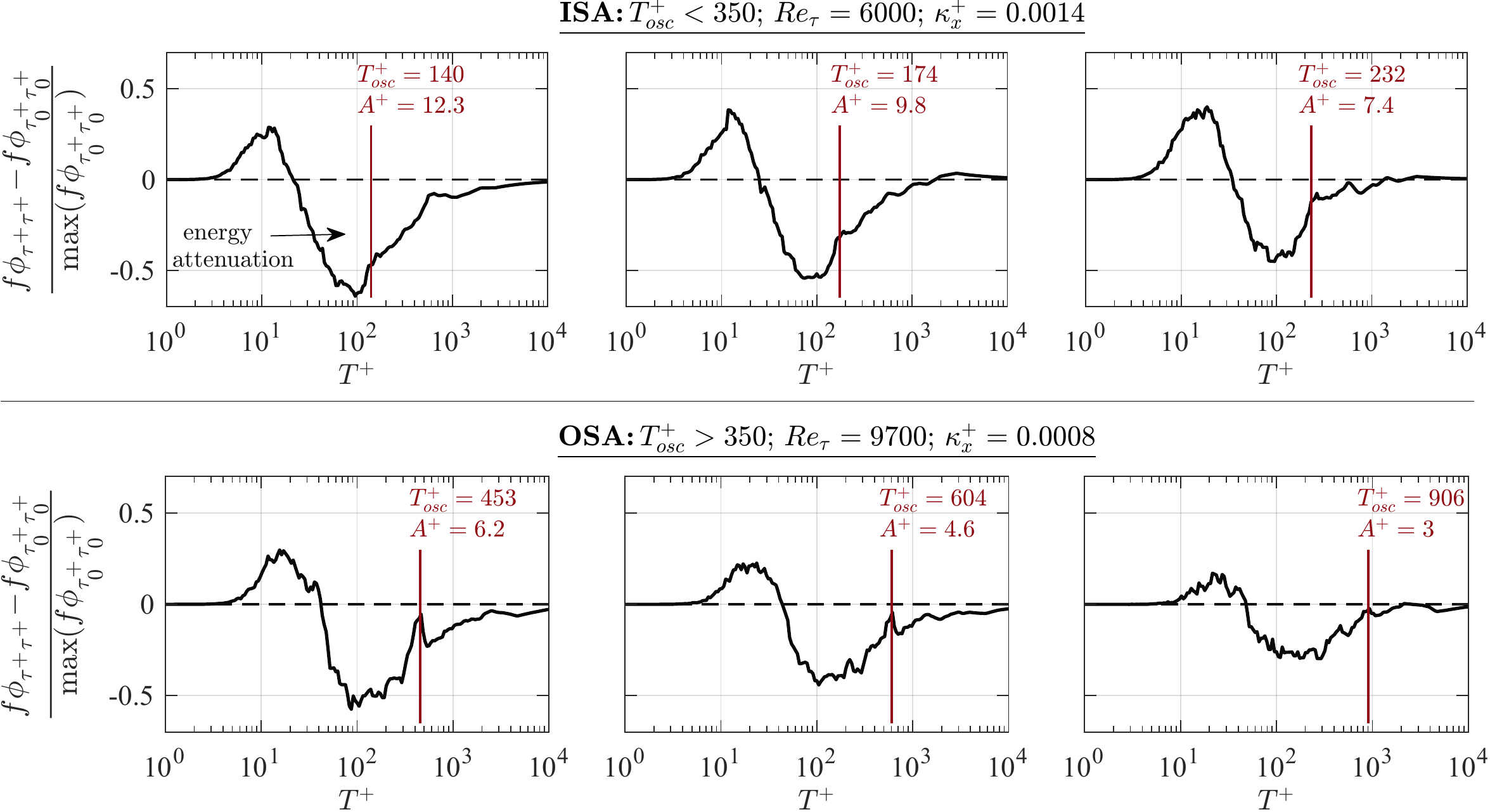} 
	\caption{Difference between the actuated and the corresponding non-actuated (reference) $\tau_w-$spectra at $Re_\tau = 6000$ (inner-scaled actuation with $T_{osc}^+ \lesssim 350$) and $Re_\tau = 9700$ (outer-scaled actuation with $T_{osc}^+ \gtrsim 350$). The negative regions represent the energy attenuated due to the actuation, and vice versa. The reference $u_{\tau_0}$ is used for the normalization. The plots demonstrate that ISA and OSA both affect a broad range of scales.}  
	\label{fig:tau_spec_diff}
\end{figure}

The $u$-spectra and $uv$ co-spectra results indicate that a broad range of turbulent scales is affected, in both the ISA and OSA cases. To investigate this further, we compute the spectra of wall shear stress fluctuations obtained using hot-wire measurements. Figure \ref{fig:tau_spec_diff} shows the difference between the actuated and the corresponding non-actuated cases. The negative regions represent the scales where energy is attenuated. The actuation time-scales are highlighted by the red line. Regardless of the time-scale of actuation, the effect of actuation is not localised to $T^+ = T_{osc}^+$ but it is felt across a broad range of scales. That is, the ISA strategy targeting the near-wall motions with $T_{osc}^+=140$ reduces the turbulent energy in the scale-range $20 \lesssim T^+ \lesssim 10^3$. Similarly, an OSA strategy targeting the larger, outer-scale motions with $T_{osc}^+ =906$ significantly impacts the small, near-wall motions at $T^+ \approx 100$. 

These observations suggest a complex inter-scale interaction between the (viscous) inner-scaled and the (inertial) outer-scaled motions. 
\dc{In a recent study, \cite{deshpande2022relationship} used the present experimental dataset to investigate this inter-scale interaction as a plausible mechanism for achieving energy-efficient drag reduction. Their analysis revealed that, for both ISA and OSA pathways, an increase in $DR$ is associated with an enhanced coupling between the inner and outer scales. This coupling is reflected in the inter scale phase relationship, wherein the inner and outer scales are found to be more `in-phase' with increasing $DR$. 
At higher Reynolds numbers, when the attached eddies and very-large-scale motions emerge and contribute meaningfully to the drag composition of the flow, this inter-scale coupling is known to be significant \citep{mathis2013estimating,deshpande2022relationship}. 
The ability to leverage this coupling, combined with its relatively low input power requirements, makes OSA a promising candidate for energy-efficient drag reduction at high Reynolds numbers \citep{marusic2021nature}.}

\section{Summary and conclusions}
We have reported on the efficacy of upstream travelling waves of spanwise oscillations as an active drag reduction strategy for high-Reynolds number turbulent flows. In Part 1, we described a numerical study of this control strategy at $Re_\tau = 950$ and $4000$. Here in Part 2, we studied its performance experimentally over a range of Reynolds numbers an order of magnitude higher than that previously investigated in the literature. At these high Reynolds numbers, the drag composition is no longer dominated by the high-frequency turbulent motions that are universally encountered very close to the wall, and it includes a significant contribution from the inertial low-frequency motions that are centered in the logarithmic region and have a signature at the wall. The relative contribution of these outer-scaled motions to skin-friction drag increases nominally as $\mathrm{ln}(Re_\tau)$ and is here found to be about 20\% at $Re_\tau \sim 10^4$. 

Following \cite{marusic2021nature}, we pursued two pathways to drag reduction: inner-scaled actuation (ISA) that targets the universal inner-scaled, near-wall features with characteristic time-scales of $T^+ \lesssim 350$, and outer-scaled actuation (OSA) that targets the inertial, outer-scaled high-$Re_\tau$ features with time-scales of $T^+ \gtrsim 350$. 

When the flow is actuated with ISA parameters ($81 \leq T_{osc}^+ \le 348$ and $6.5 \leq A^+ \le 16.3$), we find substantial drag reduction up to a maximum of $DR \approx 25\%$ at $Re_\tau =6000$. Consistent with \cite{marusic2021nature}, these results imply that spanwise wall oscillation strategies designed to target the near-wall structures are effective for reducing drag at higher Reynolds numbers, as predicted by the formulation of \cite{gatti2016reynolds}.
We note, however, that ISA mostly incurred negative net power savings between $-85\% \le NPS \le +4\%$ for the range of parameters evaluated here. On the other hand, following the OSA pathway, characterized by much lower actuation frequencies ($362 \leq T_{osc}^+ \le 1975$) and relatively low-amplitudes ($1.5 \leq A^+ \leq 7.8$), we demonstrated that considerable drag reduction could still be achieved, contrary to previous predictions (\citealp{ricco2021review}), and with much lower input power. Actuation under OSA reduced drag in the range $5\% \le DR \le 15\%$ but consistently resulted in positive net power savings with $5\% \le NPS \le 10\%$, across a wide range of Reynolds numbers ($4500 \leq Re_\tau \leq 15000$). For the current study, however, the actuation parameters are inter-dependent as $A^+ \propto \, Re_\tau/T_{osc}^+$. 
This limited the large$-T_{osc}$ OSA strategy to relatively lower amplitudes of $A^+\le 7.8$. 
The relationship also resulted in $DR$ having a nearly $Re_\tau-$independent functional form with $A^+$, with $DR$ nominally increasing linearly with $A^+$ up to $A^+ = 12$. 

For the drag reduced cases, the boundary layer is modified by the thickening of the viscous sublayer as observed in the profiles of mean streamwise velocity and a log-law diagnostic function, $y^*\, \mathrm{d}U^*/\mathrm{d}y^*$. The reduction in mean turbulent drag is accompanied by attenuation of the Reynolds normal and shear stresses and the production of turbulent kinetic energy.  Both ISA and OSA impact the boundary layer close to the wall and in the outer region.  Although the attenuation is most significant around the near-wall peak ($y^+ \approx 15$), the impact  extends up to the end of the logarithmic region. This attenuation of turbulent intensities, under both ISA and OSA pathways, is associated with the  attenuation of a broad spectrum of turbulent time scales $\mathcal{O}(10) < T^+ < \mathcal{O}(10^3)$, i.e., the low-frequency OSA strategy ($T_{osc}^+ \gtrsim 350$) reduces drag by also attenuating the high-frequency, near-wall cycle ($T^+ \approx 100$).
These findings suggest that inter-scale interactions may play a central role in the ability of OSA to considerably reduce drag  at high Reynolds numbers. 
These inter-scale interactions therefore warrant closer inspection.

\section*{Acknowledgments}
The research was funded through the Deep Science Fund of Intellectual Ventures. We thank the Scientific Instrument Manufacturing and Test Group at Intellectual Ventures Laboratory for the fabrication of SATB, Mr.\ Geoff Duke for the technical support in the laboratory, Drs.\ Kevin, Charitha de Silva and Stuart Cameron for providing helpful suggestions for the PIV experiments and Dr. Rahul Deshpande for feedback on the manuscript.

\dc{Very sadly, one of the coauthors of this work, David Wine, passed away during its preparation. He was a major contributor in realising the SATB facility and will be greatly missed.}\\

\noindent \textbf{Declaration of interests.} The authors report no conflict of interest.

\appendix
\section{Turbulence statistics scaled by actual $u_\tau$}\label{sec:appendix1}

\begin{figure}
	\centering
	\includegraphics[width=1\textwidth]{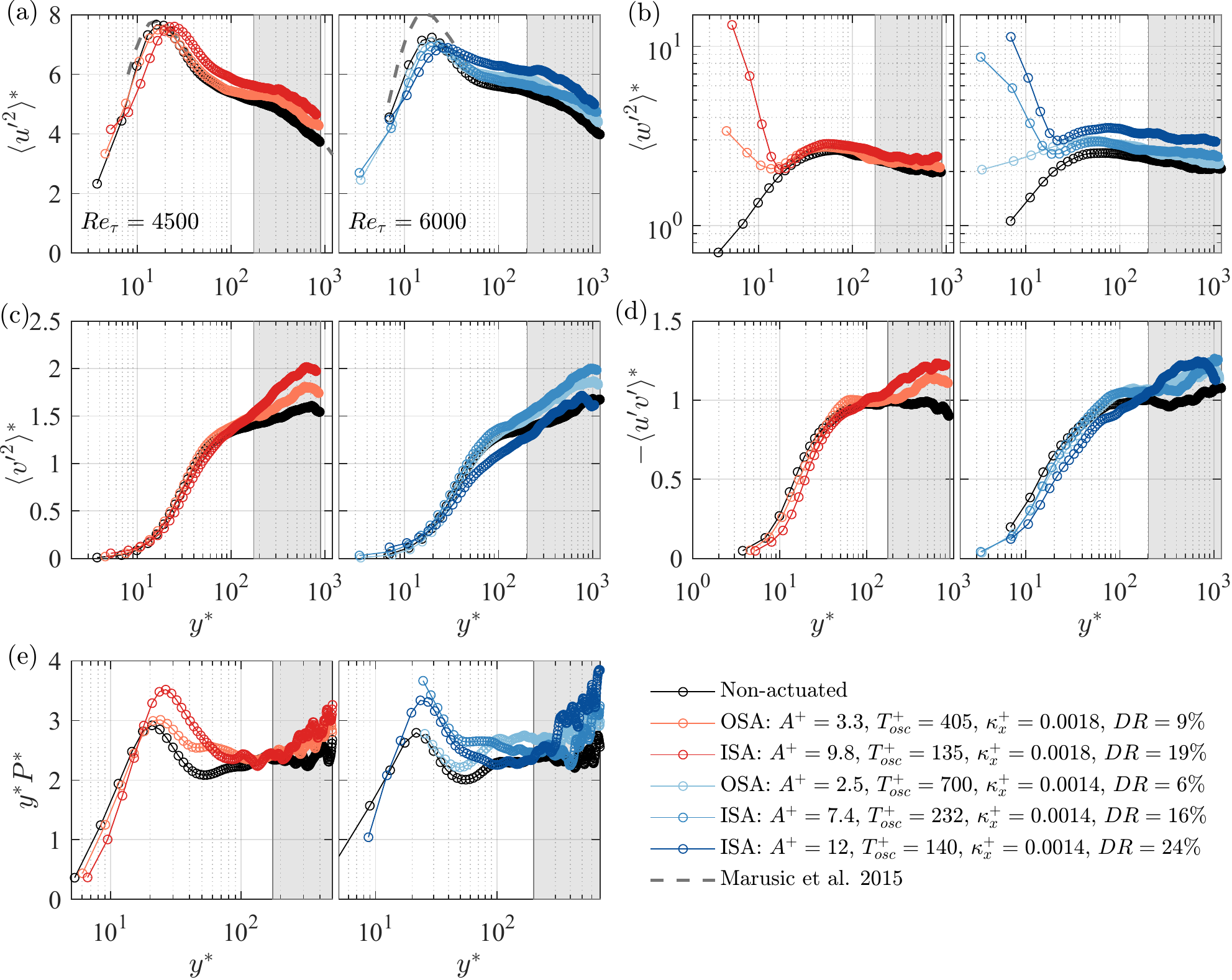}
	\caption{Turbulence statistics normalized using the actual $u_{\tau}$. (a-c) Normal stresses, (d) Reynolds shear stress and (e) premultiplied turbulence production for the non-actuated and actuated cases at $Re_\tau =4500$ and 6000. The $\langle {u^\prime}^2 \rangle^*$ profiles from the hot-wire measurements of \cite{marusic2015evolution} at matched $Re_\tau$ are included in (a) as reference for the non-actuated cases.}  
	\label{fig:appen1}
\end{figure}

In \S \ref{Sec:TurbStat}, discussion of turbulence statistics computed using the PIV data was based on the statistics being scaled by the reference non-actuated $u_{\tau_0}$ (`$+$' superscript). This scaling highlighted the absolute attenuation in the turbulence statistics for the drag-reduced cases (figures \ref{fig:PIV_stress} and \ref{fig:PIV_prod}). An alternate `inner-scaling' is generally adopted to investigate the degree of universality of the turbulence statistics where the statistics are normalized with the actual $u_{\tau}$ (`$*$' superscript). This scaling is shown here as figure \ref{fig:appen1}. 
Here, we note that $\delta^*$ ($= \delta^+\sqrt{1-DR}$) for the drag reduced cases would be lower than that of the corresponding non-actuated case. Similarly, the spatial resolution (in terms of $u_\tau$) of the PIV data for the drag-reduced cases would be correspondingly slightly higher than that of the non-actuated case.

For the prescribed inner-scaling, we see from figure \ref{fig:appen1} that the Reynolds stress profiles of the actuated cases shift closer to the non-actuated profile. For example, the magnitude of the peaks in the profiles of $\langle {u^\prime}^2 \rangle^*$ appear to be roughly similar, even though still being shifted slightly to higher wall-heights. A similar trend was observed with the LES data in Part 1 (see figure 5 in Part 1). However, we do not see the profiles collapsing away from the wall and towards the logarithmic region (shaded in grey). Instead, they seem to be systematically shifted upwards with increasing $DR$. This trend could be expected because when scaled by reference $u_{\tau_0}$ (figure \ref{fig:PIV_stress}), the profiles were observed to start merging towards the end of the logarithmic region, indicating that the absolute effects of wall-actuation were minimal at those higher wall locations. Figure \ref{fig:appen1}(e) show the premultiplied turbulence production $y^*P^*$ and its trend with $DR$. This trend is observed to be consistent with similar profiles from the LES as shown in figure 5(f) in Part 1.

\section{\dc{Photographs of the SATB}}\label{sec:appendix2}
\begin{figure}
	\begin{center}
	\subfigure[]{
	    \includegraphics[width=0.8\linewidth,trim = 0mm 10mm 0mm 10mm, clip]{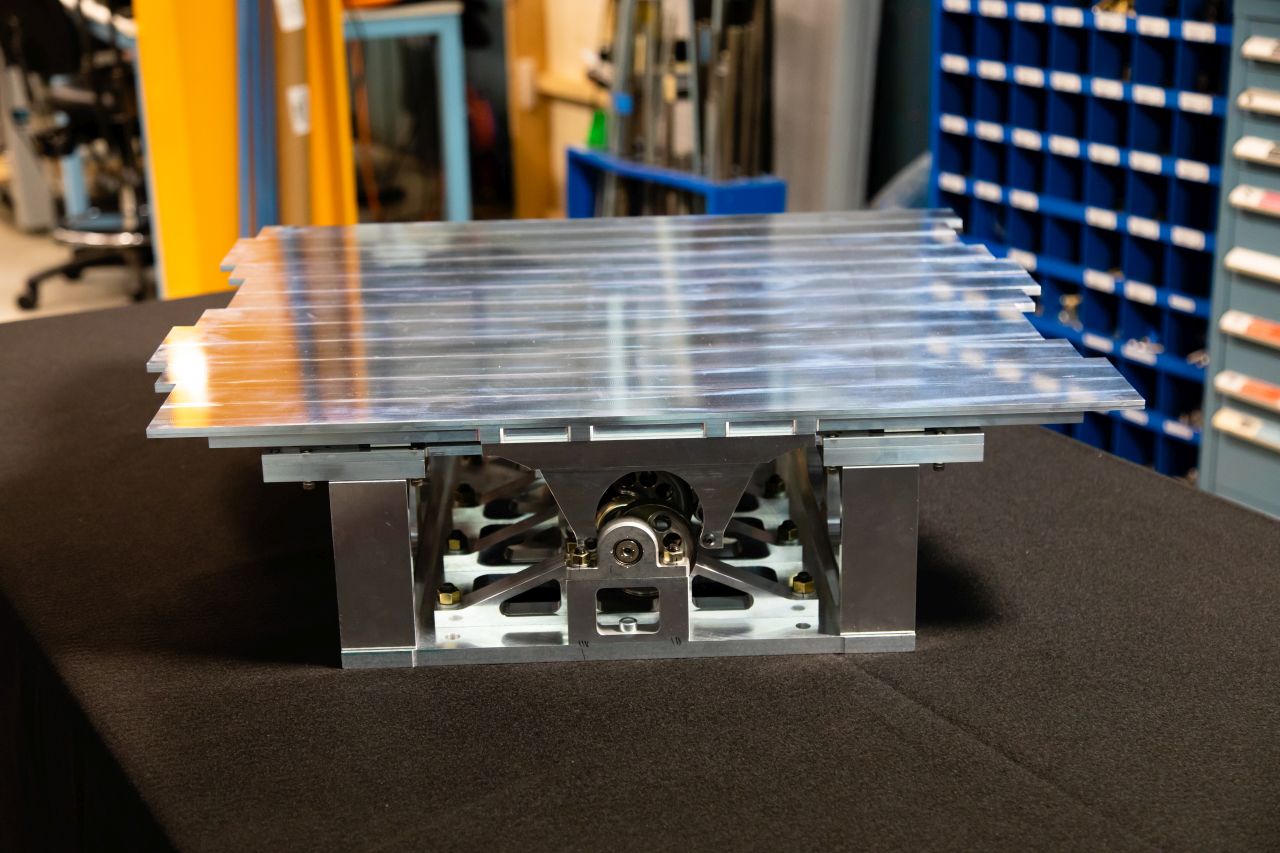}}
    \subfigure[]{
	    \includegraphics[width=0.8\linewidth,trim = 53mm 0mm 53mm 0mm, clip]{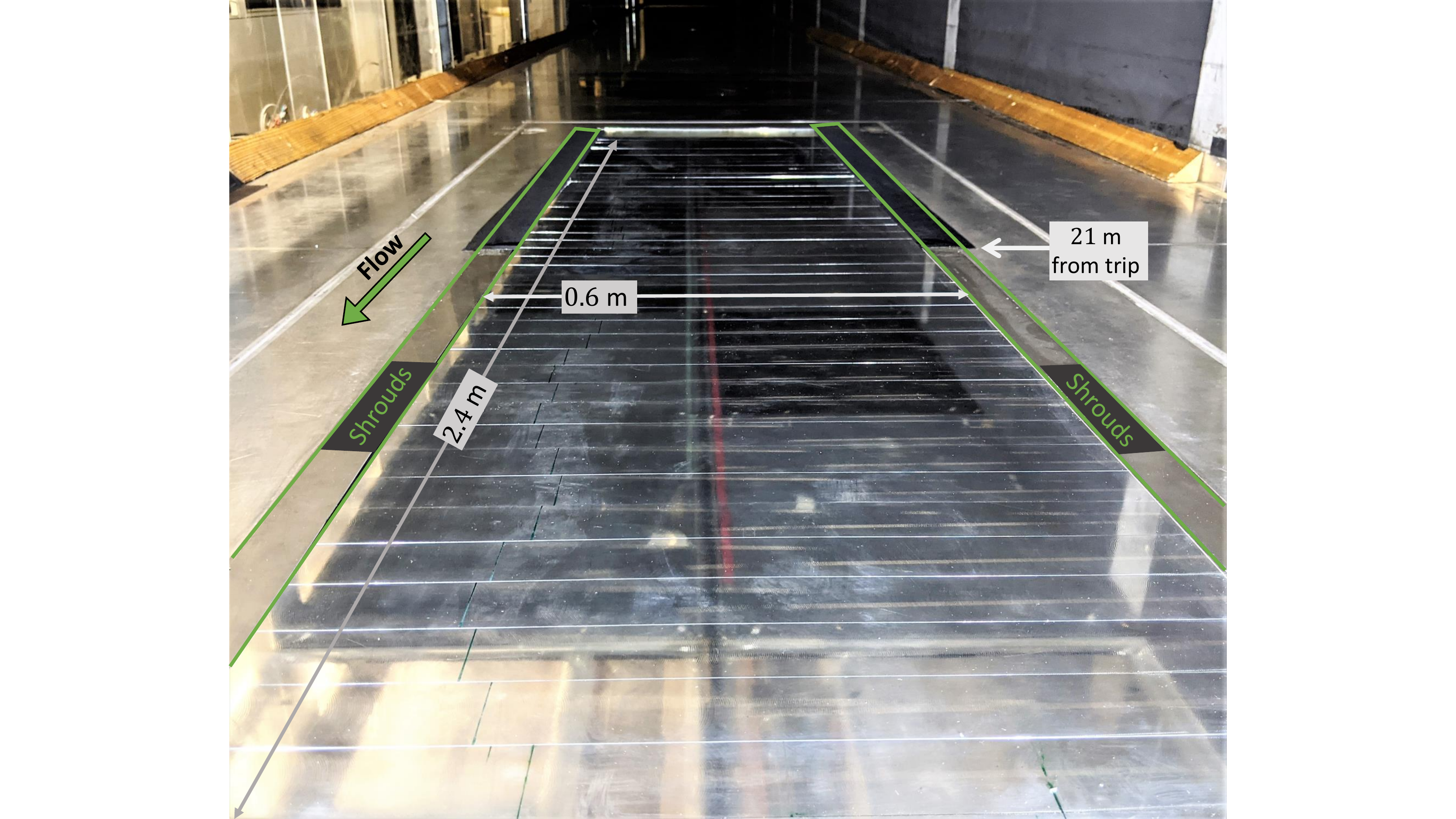}}
	\end{center}
	\caption{\dc{Photographs of SATB. (a) One of the four independently controllable $2 \lambda$ machine outside the wind tunnel and (b) all four machines installed in the wind tunnel that generate an $8 \lambda$ long streamwise travelling wave.}} 
\label{LFAT_pic}
\end{figure}

\dc{Figure \ref{LFAT_pic}(a) shows the photograph of one of the four surface actuation test bed (SATB) machines comprising of 12 slats. The phase of each slat is set by the central camshaft and the discretised streamwise travelling wave generated by this machine, of wavelength $2 \lambda$, can be seen in the slat displacement at the edge of the machine. Figure \ref{LFAT_pic}(b) shows the photograph of the SATB in the University of Melbourne wind tunnel. The four independent SATB machines are phase-synchronised to generate a $8 \lambda$ ($2.4$ m) long streamwise travelling wave.}

\bibliographystyle{jfm}
\bibliography{Paper2_FINAL}



\end{document}